\documentclass[9pt,twocolumn,twoside]{pnas-new}
\usepackage{amsmath}

\templatetype{pnasresearcharticle}
\setboolean{displaywatermark}{false}

\title{Mechanical Theory of Nonequilibrium Coexistence and Motility-Induced Phase Separation}

\author[a,b,1,2]{Ahmad K. Omar}
\author[c,1]{Hyeongjoo Row} 
\author[d,1]{Stewart A. Mallory} 
\author[c,2]{John F. Brady}

\affil[a]{Department of Materials Science and Engineering, University of California, Berkeley, CA 94720}
\affil[b]{Materials Sciences Division, Lawrence Berkeley National Laboratory, Berkeley, CA 94720}
\affil[c]{Division of Chemistry and Chemical Engineering, California Institute of Technology, Pasadena, CA 91125}
\affil[d]{Department of Chemistry, The Pennsylvania State University, University Park, PA 16802}
\leadauthor{Omar} 

\significancestatement{Phase separation, the coexistence between distinct macroscopic phases (such as oil coexisting with water), is ubiquitous in everyday life and motivated the development of the equilibrium theory of coexistence by  
Maxwell, van der Waals, and Gibbs.
However, phase separation is increasingly observed in both synthetic and living  \textit{nonequilibrium} systems, where thermodynamic principles are strictly inapplicable. 
Here, we develop a mechanical description of phase separation, offering a route for constructing phase diagrams without presuming equilibrium Boltzmann statistics.
We highlight the utility of our approach by developing a first-principles theory for motility-induced phase separation and the uniquely nonequilibrium interfacial phenomena that accompany this transition.}

\authorcontributions{Author contributions: A.K.O., H.R., S.A.M., and J.F.B. designed research, performed research, analyzed data, and wrote the paper.}
\authordeclaration{The authors declare no conflict of interest.}
\equalauthors{\textsuperscript{1}A.K.O., H.R., and S.A.M. contributed equally to this work.}
\correspondingauthor{\textsuperscript{2}To whom correspondence should be addressed.\newline E-mail: aomar@berkeley.edu or jfbrady@caltech.edu.}


\begin{abstract}
Nonequilibrium phase transitions are routinely observed in both natural and synthetic systems.
The ubiquity of these transitions highlights the conspicuous absence of a general theory of phase coexistence that is broadly applicable to both nonequilibrium and equilibrium systems. 
Here, we present a general mechanical theory for phase separation rooted in ideas explored nearly a half-century ago in the study of inhomogeneous fluids.
The core idea is that the mechanical forces within the interface separating two coexisting phases uniquely determine coexistence criteria, regardless of whether a system is in equilibrium or not.
We demonstrate the power and utility of this theory by applying it to active Brownian particles, predicting a quantitative phase diagram for motility-induced phase separation in both two and three dimensions.
This formulation additionally allows for the prediction of novel interfacial phenomena, such as an increasing interface width while moving deeper into the two-phase region, a uniquely nonequilibrium effect confirmed by computer simulations.
The self-consistent determination of bulk phase behavior and interfacial phenomena offered by this mechanical perspective provide a concrete path forward towards a general theory for nonequilibrium phase transitions. 
\end{abstract}

\begin{document}

\maketitle
\thispagestyle{firststyle}
\ifthenelse{\boolean{shortarticle}}{\ifthenelse{\boolean{singlecolumn}}{\abscontentformatted}{\abscontent}}{}

\dropcap{T}he diversity of phase behavior and pattern formation found in far-from-equilibrium systems has brought renewed focus to the theory of nonequilibrium phase transitions.
Intracellular phase separation resulting in membraneless organelles~\cite{Berry2018, Lee2020} and pattern formation on cell surfaces~\cite{Radja2019} are just a few instances in which nonequilibrium phase transitions are implicated in biological function.
Colloids~\cite{Besseling2010} and polymers~\cite{Helfand1989, Fielding2003, Wagner2010, Omar2017} subject to boundary-driven flow can experience shear-induced phase transitions and patterns that profoundly alter their transport properties.
Microscopic self-driven particles, such as catalytic Janus particles, motile bacteria, or field-directed synthetic colloids, exhibit phase transitions eerily similar to equilibrium fluids despite the absence of traditional equilibrium driving forces~\cite{Cates2015, Ivlev2015, Klymko2016, Han2017, delJunco2018, Fruchart2020}.

A general predictive framework for constructing phase diagrams for these driven systems is notably absent. 
For equilibrium systems, the formulation of a theory for phase coexistence was among the earliest accomplishments in thermodynamics.
Maxwell~\cite{Maxwell1875}, building on the work of van der Waals, derived what are now familiar criteria for phase equilibria for a one-component system: equality of temperature, chemical potential, and pressure.
These criteria are rooted in the fundamental equilibrium requirements that free energy be extensive and convex for any unconstrained degrees of freedom within a system.
The lack of such a variational principle for nonequilibrium systems has limited the theoretical description of out-of-equilibrium phase transitions.

The absence of a general theory for nonequilibrium coexistence has been particularly evident in the field of active matter.
The phenomena of motility-induced phase separation (MIPS) -- the occurrence of liquid-gas phase separation among repulsive active Brownian particles (ABPs) -- has motivated a variety of perspectives~\cite{Fily2012, Redner2013, Wittkowski2014, Takatori2015, Chakraborti2016, Solon2018, Solon2018a, Paliwal2018, Hermann2019, Hermann2021, Speck2021} in pursuit of a theory for active coexistence.
These perspectives range from kinetic models~\cite{Redner2016}, continuum and generalized Cahn-Hilliard approaches~\cite{Fily2012, Speck2014, Wittkowski2014}, large deviation theory~\cite{Whitelam2018, GrandPre2021}, and power functional theory~\cite{Hermann2019, Hermann2021}.
Some of these approaches appeal to equilibrium notions such as free energy and chemical potential~\cite{Takatori2015}, concepts which lack a rigorous basis for active systems.
Without a first-principles nonequilibrium coexistence theory, one cannot compare or assess the various perspectives. 
Despite the significant progress, a closed-form theory for the coexistence criteria for MIPS, which makes \textit{no appeals to equilibrium ideas}, remains an outstanding challenge in the field. 

Mechanics is a natural choice for describing the behavior of both equilibrium and nonequilibrium systems as it is agnostic to the underlying distribution of microstates. 
In this Article, we construct an entirely mechanical description of liquid-gas coexistence, relying only on notions such as forces and stresses.
This formulation is an extension of the mechanical perspective developed decades ago to describe coexistence and interfacial phenomena for \textit{equilibrium} systems~\cite{Davis1982, Aifantis1983a, Aifantis1983b}.
We highlight the utility of this framework by developing a theory for the coexistence criteria of MIPS and comparing our theory's predictions to results from computer simulation.
Our formulation further allows for the prediction of novel nonequilibrium interfacial behavior, such as a nonmonotonic interfacial width, as the system is taken deeper into the coexistence region. 

\section*{The Mechanics of Nonequilibrium Coexistence}
We briefly review the thermodynamics of phase separation for a one-component system undergoing a liquid-gas phase transition.
The order parameter distinguishing the liquid and gas phases is the number density $\rho\equiv N/V$ where $N$ and $V$ are the number of particles and volume, respectively. 
For simple substances at a uniform temperature $T$ below a critical temperature $T_c$, the mean-field Helmholtz free energy $\mathcal{F}(N,V,T)$ becomes concave for a range of densities, in violation of thermodynamic stability.
The system resolves this instability by separating into coexisting macroscopic domains of liquid and gas with densities $\rho^{\rm liq}$ and $\rho^{\rm gas}$, respectively.
The free energy of the phase separated system (neglecting interfacial free energy) is now $V^{\rm liq}f(\rho^{\rm liq}, T) + V^{\rm gas}f(\rho^{\rm gas}, T)$ where we have defined the free energy density $f(\rho,T) \equiv \mathcal{F}(N,V,T)/V$.
The volumes occupied by the liquid ($V^{\rm liq})$ and gas ($V^{\rm gas}$) phases sum to the total system volume $V$. 
We now obtain the coexistence criteria by minimizing the total free energy with respect to $\rho^{\rm liq}$ and $\rho^{\rm gas}$ subject to the conservation of particle number constraint (i.e., $V^{\rm liq}\rho^{\rm liq} + V^{\rm gas}\rho^{\rm gas} = V\rho$). 
This results in the familiar coexistence criteria:
\begin{subequations}
\begin{equation}
    \label{eq:equilcriteria}
    \begin{aligned}
        \mu(\rho^{\rm liq}, T) = \mu(\rho^{\rm gas}, T) = \mu^{\rm coexist}(T)\ , \\
        p(\rho^{\rm liq}, T) = p(\rho^{\rm gas}, T) = p^{\rm coexist}(T)\ ,
    \end{aligned} 
\end{equation}
where $\mu(\rho, T) = \partial f(\rho, T)/\partial \rho$ is the chemical potential, $p(\rho, T) = -f(\rho, T) + \rho \mu(\rho, T)$ is the pressure, and $\mu^{\rm coexist}(T)$ and  $p^{\rm coexist}(T)$ are the coexistence values for the chemical potential and pressure, respectively, at the temperature of interest.
It is straightforward to show that Eq.~(\ref{eq:equilcriteria}) can be equivalently expressed as:
\begin{equation}
    \label{eq:equilcriteria1}
    \begin{aligned}
    \mu(\rho^{\rm liq}) = \mu(\rho^{\rm gas}) =  \mu^{\rm coexist} \ ,\\
    \int_{\rho^{\rm gas}}^{\rho^{\rm liq}} \left[\mu(\rho) -  \mu^{\rm coexist}\right]\ d\rho = 0 \ ,
    \end{aligned} 
\end{equation}
or similarly:
\begin{equation}
    \label{eq:equilcriteria2}
    \begin{aligned}
    p(\upsilon^{\rm liq}) = p(\upsilon^{\rm gas})  = p^{\rm coexist} \ , \\
    \int_{\upsilon^{\rm gas}}^{\upsilon^{\rm liq}} \left[ p(\upsilon)  - p^{\rm coexist} \right] \ d \upsilon= 0 \ ,
    \end{aligned} 
\end{equation}
\end{subequations}
where we have defined the inverse density $\upsilon \equiv 1/\rho$ and have dropped the dependence on $T$ in Eqs.~(\ref{eq:equilcriteria1})~and~(\ref{eq:equilcriteria2}) for convenience.

The integral expressions in Eqs.~(\ref{eq:equilcriteria1})~and~(\ref{eq:equilcriteria2}) are often referred to as equal-area or Maxwell constructions~\cite{Maxwell1875} in the $\mu-\rho$ and $p-\upsilon$ planes, respectively. 
These expressions are equivalent to Eq.~(\ref{eq:equilcriteria}) and can be used to compute the coexistence curve or binodal as a function of $T$.
The spinodal boundaries enclose the region of the phase diagram in which thermodynamic stability is violated, i.e., $(\partial^2 f /\partial \rho^2)_T < 0$ or equivalently when $(\partial p /\partial \rho)_T < 0$ or $(\partial \mu /\partial \rho)_T < 0$. 
These boundaries can thus be determined by finding the densities at which $(\partial p /\partial \rho)_T = 0$ or $(\partial \mu /\partial \rho)_T = 0$ for a specified temperature.

Interestingly, the coexistence criteria presented in Eq.~(\ref{eq:equilcriteria2}) contains only the mechanical equation-of-state, a quantity which is readily defined for nonequilibrium systems (unlike, for example, chemical potential).
In fact, Eq.~(\ref{eq:equilcriteria2}) has been used in previous studies~\cite{Takatori2015, Zhang2021} to obtain the phase diagram of active systems.
However, its validity for nonequilibrium systems is questionable as its origins are clearly rooted in a variational principle that only holds for equilibrium systems. 

We are now poised to construct a theory of coexistence based purely on mechanics.
As previously noted, the order parameter for liquid-gas phase separation is density. The evolution equation for the order parameter is therefore simply the continuity equation:
\begin{equation}
\label{eq:continuity}
\frac{\partial \rho}{\partial t} + \boldsymbol{\nabla} \cdot \mathbf{j}^\rho = 0 \ ,
\end{equation}
where we are now considering a density field $\rho(\mathbf{x}; t)$ that is continuous in spatial position $\mathbf{x}$ (with $\boldsymbol{\nabla} = \partial / \partial \mathbf{x}$) and $\mathbf{j}^\rho(\mathbf{x}; t)$ is the number density flux.
A constitutive equation for the number density flux follows directly from linear momentum conservation.
This connection can be appreciated by noting that $\mathbf{j}^\rho(\mathbf{x}; t) \equiv \rho(\mathbf{x}; t)\mathbf{u}(\mathbf{x}; t)$ (where $\mathbf{u}(\mathbf{x}; t)$ is the number average velocity of particles) and is therefore proportional to the momentum density by a factor of the particle mass $m$.
Expressing linear momentum conservation with $\mathbf{j}^\rho$ (rather than the more traditional $\mathbf{u}$):
\begin{equation}
\label{eq:linearmomentum}
\frac{\partial (m \mathbf{j}_\rho)}{\partial t} + \boldsymbol{\nabla} \cdot \left (m\mathbf{j}^\rho \mathbf{j}^\rho/\rho \right) = \boldsymbol{\nabla}\cdot \boldsymbol{\sigma} + \mathbf{b} \ ,
\end{equation}
where $\boldsymbol{\sigma}(\mathbf{x}; t)$ is the stress tensor and $\mathbf{b}(\mathbf{x}; t)$ are the body forces acting on the particles. 
In simple systems,  Eqs.~(\ref{eq:continuity}) and~(\ref{eq:linearmomentum}) may constitute a closed set of coupled equations describing the temporal and spatial evolution of the density profile. 
However, the precise form of the stresses and body forces may depend on other fields, which will require additional conservation equations to furnish a closed-set of equations.

As we are interested in scenarios in which phase separation reaches a stationary state of coexistence, the continuity equation reduces to $\boldsymbol{\nabla}\cdot \mathbf{j}^{\rho} = 0$ and linear momentum conservation is now ${\boldsymbol{\nabla}\cdot(}m\mathbf{j}^\rho \mathbf{j}^\rho/\rho ) = \boldsymbol{\nabla}\cdot \boldsymbol{\sigma} + \mathbf{b}$. 
While $\mathbf{j}^{\rho} = \mathbf{0}$ for systems in equilibrium, nonequilibrium steady-states may admit nonzero fluxes\footnote{Phase-separated nonequilibrium systems with interfaces of finite curvature (i.e., if the domain of one of the coexisting phases is of non-macroscopic spatial extent) may exhibit non-zero density fluxes~\cite{Tjhung2018}.}.
However, a phase-separated system with a planar interface will satisfy $\mathbf{j}^{\rho} = \mathbf{0}$ due to the quasi-1d geometry and no-flux boundary condition. 
We restrict our discussion to macroscopic phase separation. 
Therefore, both equilibrium and nonequilibrium systems will adopt a density flux-free state, reducing the linear momentum conservation to a static mechanical force balance:
\begin{equation}
\label{eq:statics}
\mathbf{0} = \boldsymbol{\nabla}\cdot \boldsymbol{\sigma} + \mathbf{b} \ .
\end{equation}
Equation~(\ref{eq:statics}) is the mechanical condition for liquid-gas coexistence and can be used to solve for $\rho(\mathbf{x})$ with constitutive equations for $\boldsymbol{\sigma}$ and {$\mathbf{b}$}. 
The nature of these constitutive equations will also determine if other conservation equations will be required.

Let us now demonstrate that the equilibrium coexistence criteria are recovered from this mechanical perspective. 
In principle, for any system, whether it is in or out of equilibrium, microscopic expressions for Eqs.~(\ref{eq:continuity}) and~(\ref{eq:linearmomentum}) can be obtained precisely through the $N$-body distribution function and its evolution equation.
It will later be necessary to follow such an approach to obtain stresses and body forces when considering the phase coexistence of active particles. 
However, in equilibrium, the stresses and body forces can also be obtained variationally through a free energy functional. 
Consider the following free energy functional:
\begin{equation}
\label{eq:freeenergyfunctional}
\mathcal{F}[\rho] = \int_V \left[ f  + \rho \mathcal{U}^{\rm ext} + \frac{\kappa}{2} \left| \boldsymbol{\nabla}\rho \right|^2\right] \ d\mathbf{x} \ ,
\end{equation}
where $f(\rho)$ is the mean-field free energy density, $\kappa(\rho)$ is a (positive) coefficient such that the square-gradient term penalizes density gradients~\cite{Cahn1958} and $\mathcal{U}^{\rm ext}(\mathbf{x})$ represents all externally applied potential fields.
Minimizing $\mathcal{F}[\rho]$ with respect to $\rho(\mathbf{x})$~\cite{vdw1893, Cahn1958, Yang1976} {\color{black}results, after some straightforward manipulations (see SI for details), in Eq.~(\ref{eq:statics})}, allowing us to identify the reversible stress and body forces as:
\begin{subequations}
\label{eq:equil_mechanics}
\begin{equation}
\label{eq:korteweg}
    \boldsymbol{\sigma} = - p\mathbf{I} + \left(\frac{1}{2} \frac{\partial ( \kappa \rho )}{\partial \rho} \left|\boldsymbol{\nabla} \rho \right|^2 + \kappa \rho \boldsymbol{\nabla}^2 \rho \right) \mathbf{I} 
    - \kappa \boldsymbol{\nabla}\rho \boldsymbol{\nabla}\rho \ ,
\end{equation}
\begin{equation}
\label{eq:bodyforce}
\mathbf{b}  = -\rho \boldsymbol{\nabla} \mathcal{U}^{\rm ext} \ ,
\end{equation}
\end{subequations}
where the pressure is again $p(\rho) = -f(\rho) + \rho \partial f/\partial\rho$, $\mathbf{I}$ is the second-rank identity tensor. 
Note that the gradient terms appearing in Eq.~(\ref{eq:korteweg}) are the so-called Korteweg stresses~\cite{Korteweg1904}. 
The equilibrium coexistence criteria can now be obtained from Eqs.~(\ref{eq:statics})~and~(\ref{eq:equil_mechanics}). 

Without loss of generality, we take the $z$-direction to be normal to the planar interface and neglect any external potential (i.e., $\mathbf{b} = \mathbf{0}$). 
In this case, the static force balance [Eq.~(\ref{eq:statics})] reduces to $d \sigma_{zz}/d z = 0$ where we have exploited the spatial invariance tangential to the interface. 
The stress is therefore constant across the interface resulting in:
\begin{equation}
\label{eq:const_stress_equil}
    - \sigma_{zz} = p - \frac{1}{2} \left(\frac{\partial  \kappa}{\partial \rho} \rho   - \kappa \right) \left(\frac{d\rho}{dz}\right)^2 - \kappa \rho \frac{d^2\rho}{dz^2} = C\ ,
\end{equation}
where $C$ is a to-be-determined constant.

The complete density profile $\rho(z)$ can now be determined by solving Eq.~(\ref{eq:const_stress_equil}) with the appropriate boundary conditions.
For a macroscopically phase separated system, the density profile approaches constant values $\rho^{\rm liq}$ and $\rho^{\rm gas}$ as $z \rightarrow \pm \infty$. 
In these regions of constant density, the gradient terms in Eq.~(\ref{eq:const_stress_equil}) vanish and the pressure in the two phases are equal: $p(\rho^{\rm liq}) = p(\rho^{\rm gas}) = C$.
We now recognize the constant $C$ as the coexistence pressure $p^{\rm coexist}$ and recover the first of the two expected coexistence criteria in Eq.~(\ref{eq:equilcriteria2}).
Before proceeding to the second coexistence criteria, we rearrange Eq.~(\ref{eq:const_stress_equil}):
\begin{equation}
\label{eq:const_stress_equil2}
    p(\rho) - p^{\rm coexist} =  a(\rho) \frac{d^2\rho}{dz^2} + b(\rho) \left(\frac{d\rho}{dz}\right)^2 \ ,
\end{equation}
where $a(\rho) = \kappa \rho$ and $b(\rho) =  \left[  (\partial  \kappa/\partial \rho ) \rho - \kappa \right] / 2$.
To recover the second coexistence criteria in a form similar to Eq.~(\ref{eq:equilcriteria2}), we seek to integrate Eq.~(\ref{eq:const_stress_equil2}) with a variable such that the right-hand-side vanishes. 
Aifantis and Serrin~\cite{Aifantis1983a} recognized that the gradient terms can be eliminated by multiplying Eq.~(\ref{eq:const_stress_equil2}) by a weighting function $E(\rho) d\rho/dz$, where
\begin{equation}
\label{eq:E_AifantisSerrin}
   E(\rho) = \frac{1}{a(\rho)}   \exp\left( 2 \int \frac{b(\rho)} {a(\rho)} \ d\rho \right) \ ,
\end{equation}
and spatially integrating the result across the interface.
This operation eliminates the gradient terms, resulting in a coexistence criteria purely in terms of equations-of-state:
\begin{equation}
\label{eq:Econstruction}
    \int_{\rho^{\rm gas}}^{\rho^{\rm liq}} \left[ p(\rho)  - p^{\rm coexist} \right] E(\rho)\ d \rho= 0 \ .
\end{equation}
Aifantis and Serrin further established that Eq.~(\ref{eq:Econstruction}) has a unique coexistence solution, provided $a(\rho) > 0$ and $p(\rho)$ is nonmonotonic in $\rho$~\cite{Aifantis1983a}. 

Equation~(\ref{eq:Econstruction}) is no longer an equal-area construction, but such a form can be readily obtained through a simple change of variables~\cite{Solon2018, Solon2018a} $E(\rho)\equiv \partial \mathcal{E} / \partial \rho$ resulting in:
\begin{equation}
\label{eq:genequalarea}
    \int_{\mathcal{E}^{\rm gas}}^{\mathcal{E}^{\rm liq}} \left[ p(\mathcal{E})  - p^{\rm coexist} \right] \ d \mathcal{E} = 0 \ .
\end{equation}
Equation~(\ref{eq:genequalarea}) now has the form of an an equal-area construction in the $p-\mathcal{E}$ plane. 
For the equilibrium system of interest, one finds $E(\rho) = 1/\rho^2 = \upsilon^2$ and $\mathcal{E}(\rho) = \upsilon$ (multiplicative and additive constants in $E(\rho)$ and $\mathcal{E}(\rho)$ do not affect the coexistence criteria), recovering the expected equilibrium coexistence criteria [Eq.~(\ref{eq:equilcriteria2})] from our mechanical perspective. 

{\color{black} We emphasize that, for equilibrium systems, retaining higher order gradient terms in the free energy functional would not affect the resulting coexistence criteria, i.e., $\mathcal{E}(\rho) = \upsilon$ would remain the integration variable independent of the order of truncation.
This can be verified by adding higher order terms [e.g., see Ref.~\cite{Shang2011}] to Eq.~(\ref{eq:freeenergyfunctional}) (they must be even with respect to spatial gradients to satisfy the spatial inversion symmetry of the free energy) and confirming that, for the resulting stress, integration with respect to $\mathcal{E}(\rho) = \upsilon$ also eliminates the additional higher order interfacial stress terms.
This should not be surprising as, for equilibrium systems, the coexistence criteria can be derived without referencing the interface (as done at the beginning of this section), and thus should not depend on the precise details of the interface, including the truncation order.}

We further note that in order to define the spinodal \textit{without} invoking thermodynamic stability, a linear stability analysis on Eqs.~(\ref{eq:continuity})~and~(\ref{eq:linearmomentum}) [using the reversible stress Eq.~(\ref{eq:korteweg})] can be performed to determine if small density perturbations to a homogeneous base state will grow in time. 
In doing so (see Supporting Information (SI) for details), we recover the mechanical spinodal criteria $(\partial p /\partial \rho) < 0$.
This completes our discussion of the mechanics of equilibrium coexistence and stability.

\begin{figure}
	\centering
	\includegraphics[width=.4\textwidth]{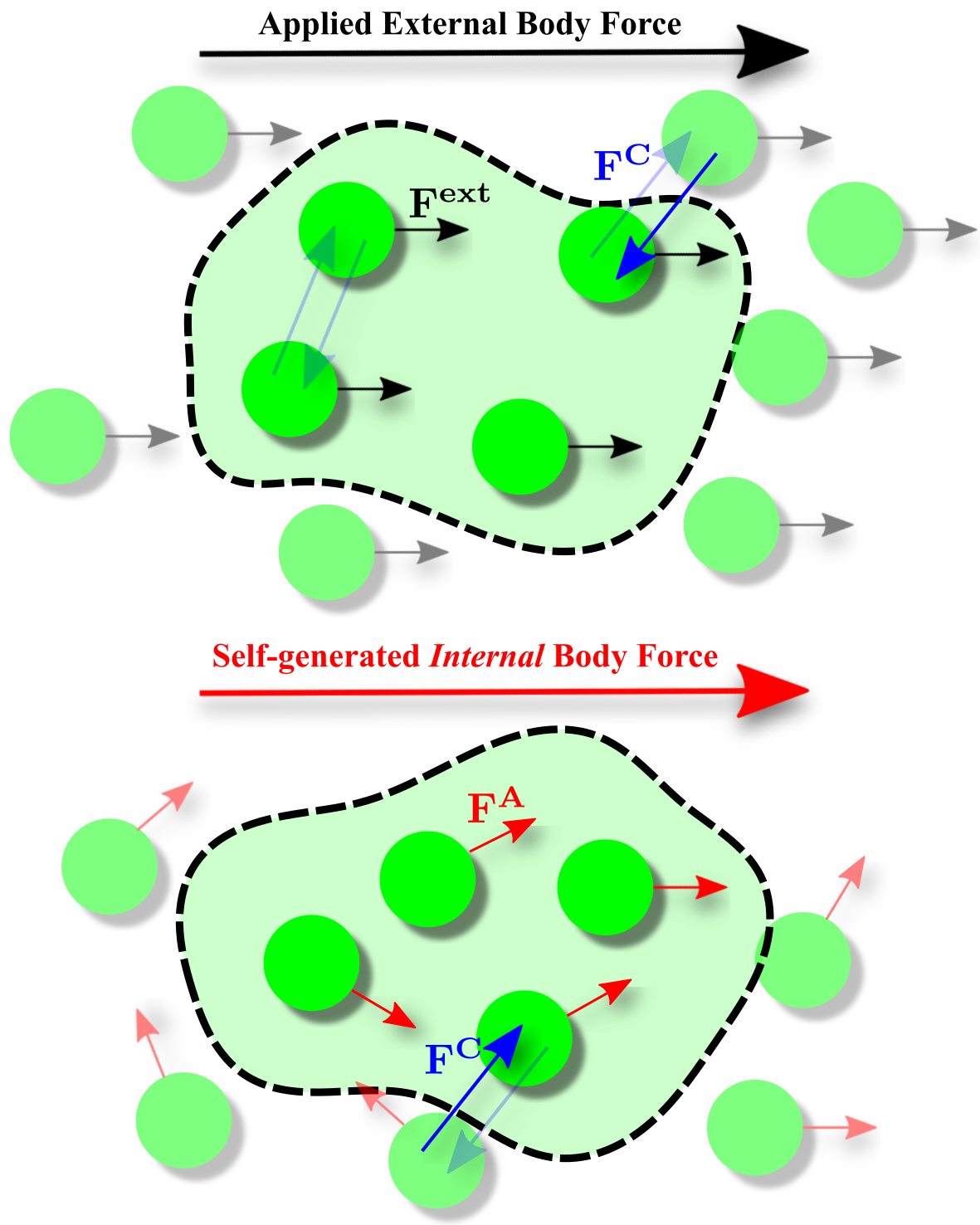}
	\caption{\protect\small{{Force balance on the particles within a control volume at steady state. Application of an external force field $\mathbf{F}^{\rm ext}$ (top) to a passive system with conservative reciprocal interaction forces $\mathbf{F}^{\rm C}$ and a system with no external forces but with active forces $\mathbf{F}^{\rm A}$ in addition to $\mathbf{F}^{C}$ (bottom).}}}
	\label{fig:figure1}
\end{figure}

For a nonequilibrium system, an additional complexity arises: the possibility of spontaneously generated \textit{internal body forces}.
The absence of applied external fields does not exclude the possibility of body forces for nonequilibrium systems. 
A general nonequilibrium coexistence criteria for liquid-gas phase separation must therefore account for these internal body forces.
To understand this physically, let us consider a steady state force balance on a collection of particles in a control volume [see Fig.~\ref{fig:figure1}].
Application of an external force field on the particles results in a net volumetric force acting on the particles: a \textit{body force}.
By Newton's third law, interparticle interactions do not give rise to a net volumetric force within the volume interior. 
It is only at the surface of the control volume that interparticle forces (exerted by particles outside the volume on the interior particles) are non vanishing, resulting in \textit{stresses}.
The polarization of active forces (see bottom of Fig.~\ref{fig:figure1}) results in a net active force within the volume, behaving similarly to an external force field~\cite{Yan2015}.

At steady state, the self-generated body force density due to nonequilibrium forces must balance a stress difference across the volume. 
In this case, the steady-state one dimensional (1d) mechanical balance is $d\sigma_{zz}/dz + b_z = 0$.
For a one dimensional system, the body force can always be expressed as $b_z = d\sigma^b/dz$ and the mechanical balance can now be expressed as $d(\sigma_{zz} + \sigma^b)/dz = 0$.
This newly defined effective stress $\Sigma \equiv \sigma_{zz} + \sigma^b$ is, just as before, constant spatially. 
Expressing $\Sigma$ as a second-order gradient expansion in density:
\begin{equation}
\label{eq:Mech_Theory_const_stress_equil}
     - \Sigma = \mathcal{P}(\rho) - a(\rho) \frac{d^2\rho}{dz^2} - b(\rho) \left(\frac{d\rho}{dz}\right)^2 = C \ ,
\end{equation}
where $\mathcal{P}(\rho)$ is a dynamic or effective pressure.
We again recognize that, as the gradients must vanish in the bulk phases, $\mathcal{P}(\rho^{\rm liq}) = \mathcal{P}(\rho^{\rm gas})= C$, where we identify the constant as the coexistence effective pressure $\mathcal{P}^{\rm coexist}$.
The second coexistence criteria can be found analogously as before through the use of an integrating factor $E(\rho) d\rho/dz$, where $E(\rho)$ is defined in Eq.~(\ref{eq:E_AifantisSerrin}). 
The two coexistence criteria are then:
\begin{subequations}
\label{eq:Mech_Theory}
\begin{equation}
\label{eq:Mech_Theory_equal_P_a}
    \mathcal{P}(\mathcal{E}^{\rm liq}) = \mathcal{P}(\mathcal{E}^{\rm gas}) = \mathcal{P}^{\rm coexist} \ ,
\end{equation}
\begin{equation}
\label{eq:Mech_Theory_gen_eqal_area_b}
    \int_{\mathcal{E}^{\rm gas}}^{\mathcal{E}^{\rm liq}} \left[ \mathcal{P}(\mathcal{E})  - \mathcal{P}^{\rm coexist} \right] \ d \mathcal{E} = 0 \ ,
\end{equation}
where 
\begin{equation}
\label{eq:equalareavariable}
   \frac{\partial\mathcal{E}}{\partial \rho} = \frac{1}{a(\rho)}   \exp\left( 2 \int \frac{b(\rho)} {a(\rho)} \ d\rho \right) \ .
\end{equation}
\end{subequations}
Equation~(\ref{eq:Mech_Theory}) \textit{is the general nonequilibrium coexistence criteria for liquid-gas phase separation}. 

{\color{black} The powerful idea that coexistence criteria can be extracted from knowledge of interfacial mechanics was, to the best of our knowledge, first proposed by Aifantis and Serrin~\cite{Aifantis1983a} in the context of equilibrium systems.
Solon and co-workers proposed a similar gradient-expansion based approach beginning with a generalized Cahn-Hilliard model~\cite{Solon2018, Solon2018a}.
The criteria derived herein [Eq.~(\ref{eq:Mech_Theory})] makes clear that for nonequilibrium phase separation, one criteria is \textit{always} equality of dynamic pressure while the other is obtained from knowledge of the interfacial stresses and body forces. 
}

Application of this criteria to determine the phase diagram will require expressing the dynamic pressure $\mathcal{P}(\rho)$ as a second order density gradient expansion in order to identify the equal-area construction variable $\mathcal{E}(\rho)$.
Furthermore, provided that a timescale exists such that this dynamic pressure can also be defined for time-dependent states, the spinodal criteria is now $(\partial \mathcal{P} /\partial \rho) < 0$, as shown in the SI. 
We now proceed to obtain the dynamic pressure of active Brownian particles and apply this nonequilibrium coexistence criteria.

\section*{The Mechanical Theory of MIPS} 
For a theoretical prediction of the phase diagram of active Brownian particles, our mechanical perspective requires expressions for the dynamic pressure, $\mathcal{P}(\rho)$, and the coefficients of the leading gradient terms, $a(\rho)$ and $b(\rho)$.
These quantities are needed to calculate the appropriate integration variable $\mathcal{E}(\rho)$ such that  Eq.~(\ref{eq:Mech_Theory}) is satisfied.
To derive these quantities, we require expressions for the stress $\boldsymbol{\sigma}$ and body forces $\mathbf{b}$ without invoking a variational principle.
These constitutive equations can be obtained systematically, beginning with the equations-of-motion describing the motion of the microscopic degrees of freedom. 
We consider active Brownian particles with overdamped translational and rotational equations-of-motion describing the position $\mathbf{r}_{\alpha}$ and orientation $\mathbf{q}_{\alpha}$ ($|\mathbf{q}_\alpha| = 1$) of particle $\alpha$ as:
\begin{subequations}
\label{eq:eqn_motion}
\begin{equation}
\label{eq:eqn_motion_trans}
\dot{\mathbf{r}}_{\alpha} = U_0 \mathbf{q}_{\alpha} + \frac{1}{\zeta} \mathbf{F}^{\rm C}_{\alpha} \ ,
\end{equation}
\begin{equation}
\label{eq:eqn_motion_rot}
\dot{\mathbf{q}}_{\alpha} = \boldsymbol{\Omega}^{R}_{\alpha} \times \mathbf{q}_{\alpha} \ ,
\end{equation}
\end{subequations}
where $\zeta$ is the translational drag coefficient, and $\mathbf{F}^{\rm C}_{\alpha}$ is the interparticle force on particle $\alpha$. 
The orientation of a particle evolves under the influence of a stochastic angular velocity $\boldsymbol{\Omega}^{R}_{\alpha}$ which follows the usual white noise statistics with a mean of $\left \langle \boldsymbol{\Omega}^{R}_{\alpha}(t)\right \rangle \!=\!\mathbf{0}$ and a variance of ${\left \langle \boldsymbol{\Omega}^{R}_{\alpha}(t) \boldsymbol{\Omega}^{R}_{\beta}(t') \right \rangle = \left(2 / \tau_R\right)  \delta_{\alpha \beta} \delta(t-t')\mathbf{I}}$ where $\tau_R$ is the reorientation time and $\delta_{\alpha \beta}$ is the Kronecker delta.
We aim to describe the strongly active (athermal) limit of hard active disks and spheres where the phase diagram for these systems are fully described by two geometric parameters: the volume (or area) fraction $\phi \equiv v_p \rho$ (where $v_p$ is the area ($d=2$) or volume ($d=3$) of a particle) and the dimensionless intrinsic run length $\ell_0/D$, where $\ell_0 \equiv U_0\tau_R$, $D$ being the particle diameter and $U_0$ is the intrinsic active speed.
We therefore choose a conservative force $\mathbf{F}^{\rm C}_{\alpha}$ that results in hard-particle interactions, as further detailed in the Materials and Methods.

The probability density $f_N(\mathbf{\Gamma}; t)$ of finding the system in a microstate $\boldsymbol{\Gamma} = (\mathbf{r}^N, \mathbf{q}^N)$ at time $t$ satisfies a conservation equation $\partial f_N / \partial t = \mathcal{L}f_N$, where $\mathcal{L}$ is the relevant dynamical operator specific to the microscopic equations-of-motion [e.g., Eq.~(\ref{eq:eqn_motion})]. 
Conservation equations needed to describe the density-field (at a minimum, the continuity equation and linear momentum conservation) can be directly obtained through this dynamical operator and distribution function.
For example, the continuity equation for the ensemble-averaged microscopic density $\rho(\mathbf{x}; t) = \left \langle \hat{\rho}(\mathbf{x}) \right \rangle  = \left \langle \sum_{\alpha = 1}^N\delta(\mathbf{x} - \mathbf{r}_{\alpha}) \right \rangle$ is given by $\partial \rho /\partial t = \int_{\gamma} \hat{\rho}\mathcal{L}f_N \ d\boldsymbol{\Gamma}$ where $\gamma$ is the phase-space volume. 
An expression for linear momentum conservation and all other required conservation equations can be similarly obtained.

In the case of ABPs, $\mathcal{L}$ is the Fokker-Planck (or Smoluchowski) operator.
For brevity, this operator and the conservation equations resulting from it are provided in the Materials and Methods and a complete derivation can be found in the SI.
Here, we only include only the necessary results to obtain the MIPS phase diagram.

The linear momentum balance for overdamped ABPs is found to simply be
$\mathbf{0} = \boldsymbol{\nabla}\cdot\boldsymbol{\sigma} + \mathbf{b}$, where the inertial terms [the left-hand-side of Eq.~(\ref{eq:linearmomentum})] are identically zero.
The stress is identified as $\boldsymbol{\sigma} = \boldsymbol{\sigma}^{\rm C}$, where $\boldsymbol{\sigma}^{\rm C}$ is the stress generated by the conservative interparticle forces. 
The body forces are given by $\mathbf{b} = -\zeta \mathbf{j}^{\rho} + \zeta U_0 \mathbf{m}$, where  $-\zeta \mathbf{j}^{\rho}$ is the drag force density and $\zeta U_0 \mathbf{m}$ is the active force density arising from the polarization density field  $\mathbf{m}(\mathbf{x}; t) = \left \langle \sum_{\alpha = 1}^N \mathbf{q}_{\alpha}\delta(\mathbf{x} - \mathbf{r}_{\alpha})\right \rangle$. 
For the quasi-1d system, the active force density is the sole body force as $\mathbf{j}^{\rho} = \mathbf{0}$, reducing the linear momentum balance to: 
\begin{equation}
\label{eq:mech_force_bal_ABP_tensorial}
    \mathbf{0} = \boldsymbol{\nabla} \cdot \boldsymbol{\sigma}^{\rm C} + \zeta U_0 \mathbf{m} \ .
\end{equation}
Activity thus manifests as a body force~\cite{Yan2015, Rodenburg2017, Epstein2019, Omar2020} rather than a true stress.

An added complexity for ABP coexistence is that we now require an additional conservation equation for the polarization density field $\mathbf{m}$ as it appears in Eq.~(\ref{eq:mech_force_bal_ABP_tensorial}). 
This is given by:
\begin{equation}
    \label{eq:m_eqn_tensor}
    \mathbf{m} = -\frac{\tau_R}{d-1}\boldsymbol{\nabla} \cdot \mathbf{j^m}\ . 
\end{equation}
The form of Eq. (\ref{eq:m_eqn_tensor}) allows us to write an \textit{effective} stress for the system as:
\begin{equation}
    \label{eq:dynamic_stress}
    \boldsymbol{\Sigma} = \boldsymbol{\sigma}^{\rm C} + \boldsymbol{\sigma}^{\rm act} \ ,
\end{equation}
where we have defined the active or ``swim''~\cite{Takatori2014} stress as $\boldsymbol{\sigma}^{\rm act} = -\zeta U_0 \tau_R \mathbf{j^m}/(d-1)$~\cite{Omar2020}.
It is important to note here that the effective stress we define here is not a true stress [just as the Maxwell stress tensor is not a true stress tensor~\cite{Rinaldi2002}].
This distinction between true stresses ($\boldsymbol{\sigma}^{\rm C}$) and effective stresses ($\boldsymbol{\Sigma}$) was found to be crucial~\cite{Omar2020} in computing the surface tension of ABPs~\cite{Bialke2015, Patch2018, Hermann2019b}, which requires the true stress tensor~\cite{Omar2020, Lauersdorf2021}.

In our derivation of the effective stress [Eq.~(\ref{eq:dynamic_stress})] we have made no approximations.
However, to utilize our nonequilibrium coexistence criteria, we must be able to express $\Sigma = \sigma^{\rm C}_{zz} + \sigma^{\rm act}_{zz}$ in terms of bulk equations-of-state and density gradients. 
A gradient expansion of the conservative interparticle stress $\sigma_{zz}^C$ results in the bulk interaction pressure $p_{\rm C}(\rho)$ and Korteweg-like terms with coefficients related to the pair-interaction potential and pair-distribution function~\cite{Yang1976}.
In the SI, we show the coefficients on the gradient terms associated with $\sigma_{zz}^C$ scale as $\zeta U_0D$ -- the stress scale for active hard-particle collisions -- while, as we demonstrate next, the gradient terms in the active stress scale as $\zeta U_0\ell_0$. 
As MIPS occurs at $\ell_0/D \gg 1$, we can safely discard the Korteweg-like terms and approximate the conservative interparticle stress as $\sigma^{\rm C}_{zz} \approx -p_{\rm C}(\rho)$.

We now turn our focus to an expression for the active stress $\sigma^{\rm act}_{zz}$ in terms of bulk equations-of-state and density gradients.
Deriving a constitutive equation for the polarization flux $\mathbf{j^m}$ results in $\sigma^{\rm act}_{zz}$ taking the following form:
\begin{equation}
     \label{eq:j_m_scalar}
     \sigma^{\rm act}_{zz} (z) =
     -\frac{\zeta \ell_0  U_0 \overline{U}(\rho)}{d(d-1)}  \left(\rho(z) + dQ_{zz}(z)\right) \ ,
 \end{equation}
where $Q_{zz}$ is the normal component of the traceless nematic density field $\mathbf{Q}(\mathbf{x}; t) = \left \langle \sum_{\alpha = 1}^N ((\mathbf{q}_{\alpha}\mathbf{q}_{\alpha} - \mathbf{I}/d )\delta(\mathbf{x} - \mathbf{r}_{\alpha})\right \rangle$. $U_0\overline{U}(\rho)$ is the density-dependent average speed of the particles. 
In the absence of interparticle interactions, the normalized speed $\overline{U}(\rho) = 1$ as particle motion is unencumbered.
An equation-of-state for $\overline{U}(\rho)$ is required to describe this bulk contribution of the active stress.
The nematic field satisfies its own conservation equation which takes the following form at steady-state:
\begin{subequations}
\label{eq:nematic}
\begin{equation}
\label{eq:nematic_a}
Q_{zz}(z) = -\frac{\tau_R}{2d}\frac{d }{dz} j^Q_{zzz} \ ,
\end{equation}
\begin{equation}
\label{eq:nematic_b}
j^Q_{zzz} = U_0 \overline{U}(\rho) B_{zzz}(z) + \left(\frac{3 \overline{U}(\rho)}{d + 2}   - \frac{1}{d}  \right) U_0 m_z(z)  + \frac{1}{d \zeta} \frac{d p_{\rm C}}{d z} \ ,
\end{equation}
\end{subequations}
where $B_{zzz}$ is the relevant component of the traceless third orientational moment ${\mathbf{B}=\left \langle \sum_{\alpha = 1}^N ((\mathbf{q}_{\alpha}\mathbf{q}_{\alpha}\mathbf{q}_{\alpha} - \boldsymbol{\alpha} \cdot \mathbf{q}_{\alpha}/(d+2) )\delta(\mathbf{x} - \mathbf{r}_{\alpha})\right \rangle}$, where $\boldsymbol{\alpha}$ is a fourth-rank isotropic tensor (see Materials and Methods or SI).
As we are interested in density gradients up to second order, we can safely close the hierarchy of orientational moments by setting  $\mathbf{B} = \mathbf{0}$.
We also recognize from linear momentum conservation Eq.~(\ref{eq:mech_force_bal_ABP_tensorial}) that $\zeta U_0 m_z - d p_{\rm C}/d z = 0$, allowing us to substitute $p_{\rm C}$ in place of $m_z$ in Eq.~(\ref{eq:nematic_b}).
Our expression for the effective stress is now:
\begin{equation}
\label{eq:mechanicaldiffeq}
 - \Sigma = p_{\rm C} + p_{\rm act} - \frac{3 \ell_0^2}{2d(d - 1)(d + 2)}  \overline{U}(\rho) \frac{d}{d z} \left( \overline{U}(\rho) \frac{d p_{\rm C}}{d z}  \right)\ ,
\end{equation}
where $p_{\rm act} = \rho\zeta \ell_0  U_0 \overline{U}(\rho)/ d(d-1)$ is the active pressure~\cite{Fily2014, Mallory2014, Takatori2014, Solon2015, Solon2015a, Epstein2019, Omar2020} --- an effective pressure emerging from the active body force density. 

The mechanical terms needed to apply our nonequilibrium coexistence criteria, for a given activity $\ell_0$,  can now be identified as:
\begin{subequations}
\begin{equation}
    \label{eq:caliP_ABP_a}
    \mathcal{P}(\rho) = p_{\rm C} + p_{\rm act}\ ,
\end{equation}
\begin{equation}
    \label{eq:a_ABP_b}
    a(\rho) = \frac{3 \ell_0^2}{2d(d - 1)(d + 2)}\overline{U}^2  
    \frac{\partial p_{\rm C}}{\partial \rho}\ .
\end{equation}
\begin{equation}
    \label{eq:b_ABP_c}
    b(\rho) = \frac{3 \ell_0^2}{2d(d - 1)(d + 2)}\overline{U}  
    \frac{\partial}{\partial \rho}\left[\overline{U}
    \frac{\partial p_{\rm C}}{\partial \rho}\right]\ ,
\end{equation}
\end{subequations}

Equations~(\ref{eq:equalareavariable}),~(\ref{eq:caliP_ABP_a}),~and~(\ref{eq:a_ABP_b}) allow us to identify $\mathcal{E}(\rho) = p_{\rm C} (\rho)$. 
The coexistence criteria for MIPS is therefore:
\begin{subequations}
\label{eq:Mech_Theory_MIPS}
\begin{equation}
\label{eq:Mech_Theory_MIPS_equal_P_a}
    \mathcal{P}(p_{\rm C}^{\rm liq}) = \mathcal{P}(p_{\rm C}^{\rm gas}) = \mathcal{P}^{\rm coexist} \ ,
\end{equation}
\begin{equation}
\label{eq:Mech_Theory_MIPS_gen_eqal_area_b}
    \int_{p_{\rm C}^{\rm gas}}^{p_{\rm C}^{\rm liq}} \left[ \mathcal{P}(p_{\rm C})  - \mathcal{P}^{\rm coexist} \right] \ d p_{\rm C} = 0 \ .
\end{equation}
\end{subequations}
Furthermore, the spinodal criteria is indeed found to be $(\partial \mathcal{P} /\partial \rho) < 0$ (see SI for details).

To apply this coexistence criteria we need to know the functional form of $p_{\rm C}(\rho, \ell_0)$ and $p_{\rm act}(\rho, \ell_0)$ (or equivalently $\overline{U}$) as a function of volume fraction $\phi$ (in place of $\rho$) and activity $\ell_0/D$.
A detailed theoretical treatment for these equations-of-state will require a theory for the pair-distribution function $g(\mathbf{r}, \mathbf{q})$ where $\mathbf{r}$ and $\mathbf{q}$ are the separation vector and relative orientation vector between particle pairs, respectively.
The description of nonequilibrium pair-correlations is an active area of investigation. 
Theories applicable in the dilute limit have been proposed~\cite{Squires2005}, and recent developments have been made towards our understanding of strongly interacting systems~\cite{Tociu2019, Tociu2022}.
{\color{black}Closure relations rooted in ideas from dynamical density functional theory~\cite{teVrugt2020} have also been proposed for a variety of active systems, including ABPs~\cite{Hermann2019, Hermann2019b},  hydrodynamically interacting microswimmers~\cite{Menzel2016}, and active rods~\cite{Bertin2015}, to name a few.}

\begin{figure}
	\centering
	\includegraphics[width=.475\textwidth]{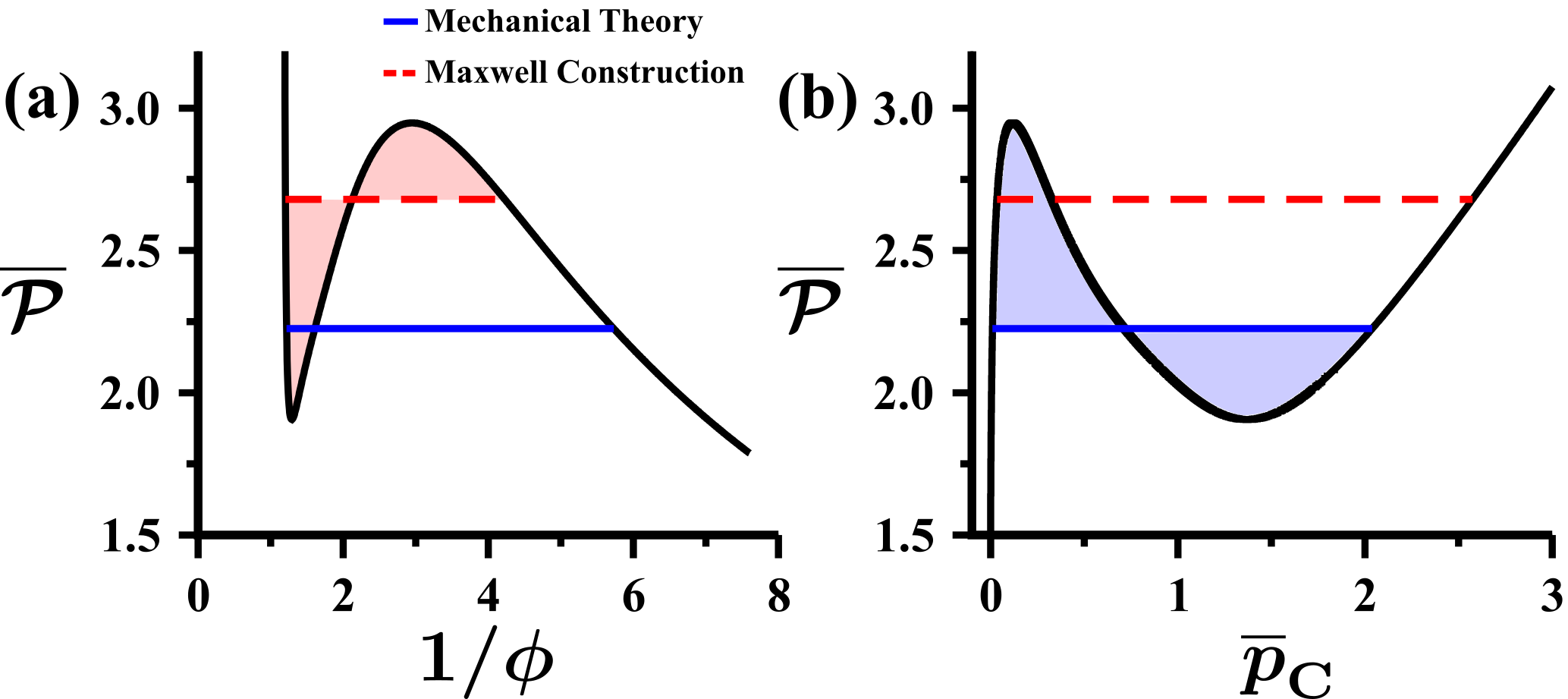}
	\caption{\protect\small{{Predicted homogeneous equation-of-state for 2d athermal ABPs~\cite{Mallory2021} with $\ell_0/D \approx 31.2$. (a) The equal-area Maxwell construction in the $\mathcal{P}-\phi^{-1}$ plane overestimates the coexistence pressure as predicted from (b) the equal-area construction in the $\mathcal{P}-p_{\rm C}$ established by our nonequilibrium theory. $\mathcal{P}$ and $p_{\rm C}$ are made dimensionless by $\zeta U_0 D/v_p$.}}}
	\label{fig:figure2}
\end{figure}

An alternative approach is to obtain these equations-of-state directly from particle-based simulations in regions of the $\phi-\ell_0$ plane where the system remains homogeneous. 
This measured behavior can then be extrapolated to regions of the $\phi-\ell_0$ plane where the equations-of-state cannot be directly obtained by leveraging a number of physical considerations (e.g., $p_{\rm C}$ is a monotonically increasing function of both $\phi$ and $\ell_0$), as detailed in Ref.~\cite{Mallory2021}.
In two dimensions (2d), we utilize the equations-of-state developed in Ref.~\cite{Mallory2021} and follow a similar procedure to develop three dimensional (3d) versions, provided in the SI. 
We note that in both 2d~\cite{Digregorio2018} and 3d~\cite{Omar2021}, ABPs can exhibit an order-disorder transition.
The theory presented here applies only to scenarios where the sole order parameter is density. 
We therefore limit our focus to polydisperse ABPs in 2d (eliminating any potential ordered phase) and, in 3d, recognize that the liquid-gas transition is metastable with respect to a fluid-crystal transition for much of the phase diagram~\cite{Omar2021}. 

\begin{figure}[t]
	\centering
	\includegraphics[width=.475\textwidth]{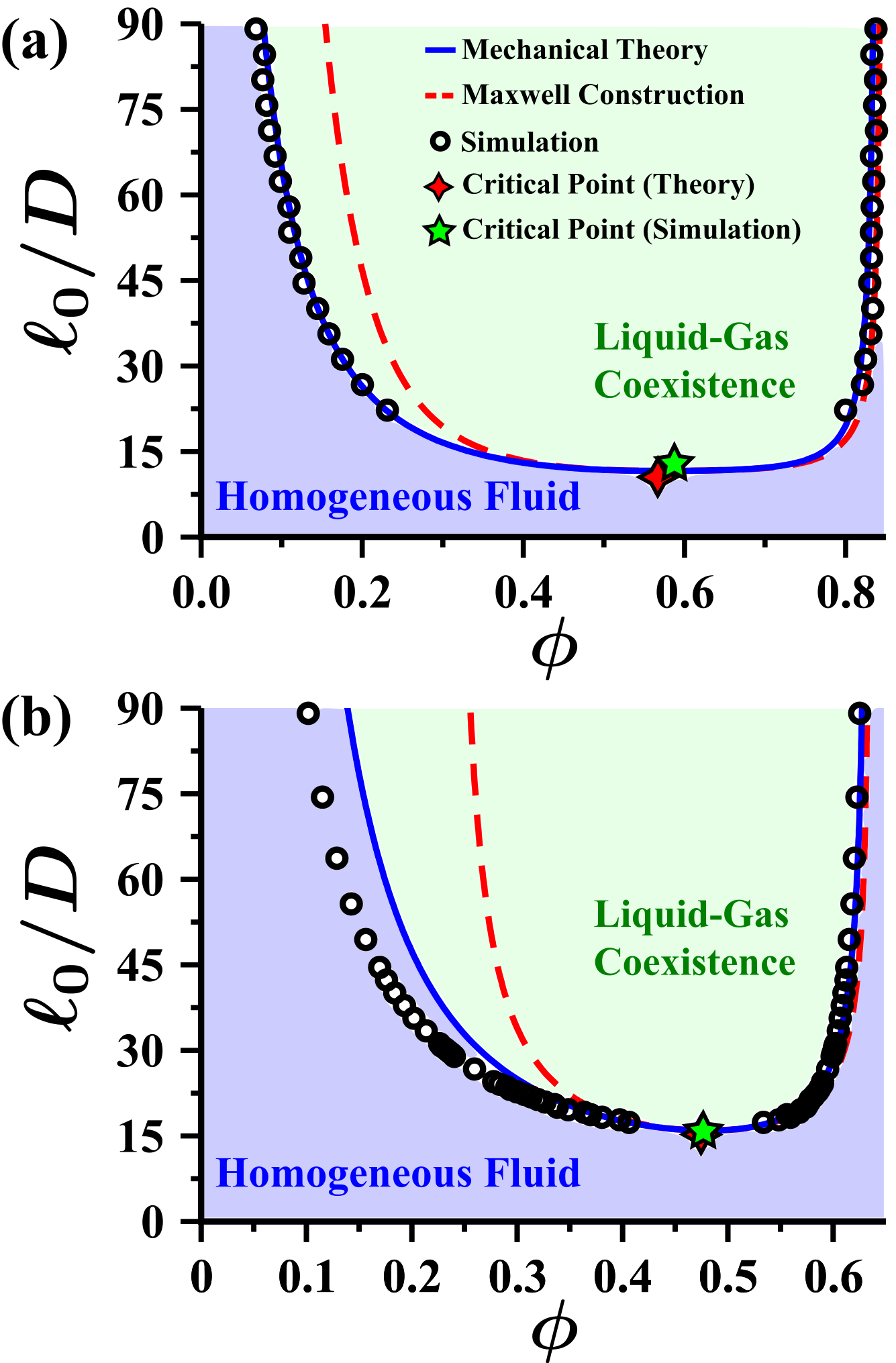}
	\caption{\protect\small{{Coexistence curves for athermal active Brownian (a) disks (2d) and (b) spheres (3d). Coexisting densities were obtained from slab simulation data collected in this work (2d) and from Ref.~\cite{Omar2021} (3d). Critical points displayed were estimated from simulations in Refs.~\cite{Siebert2018} (2d) and \cite{Omar2021} (3d). Regions of coexistence and homogeneity are shaded on the basis of our theoretical predictions.}}}
	\label{fig:figure3}
\end{figure}

Figure~\ref{fig:figure2} compares the results of performing the equal-area construction in the $\mathcal{P}-p_{\rm C}$ plane with the naive application of the Maxwell (equilibrium) equal-area-construction in the $\mathcal{P}-\upsilon$ plane (where $\upsilon \sim 1/\phi$). 
The equilibrium construction overestimates the coexistence pressure in comparison to our nonequilibrium theory, resulting in less disparate coexisting densities.
This trend holds in both two and three dimensions (see the binodals presented in Fig.~\ref{fig:figure3}) and is exacerbated with increasing activity.

We now compare our theory with extensive simulations of polydisperse hard-disks (2d) performed in this study [see Fig.~\ref{fig:figure3}(a)] and simulations of monodisperse hard-spheres (3d) conducted in Ref.~\cite{Omar2021} [see Fig.~\ref{fig:figure3}(b)].
The agreement between our theory and simulation data is nearly perfect in 2d and, while there is less agreement in 3d, the nonequilibrium theory provides a substantially improved binodal in comparison to that predicted by the equilibrium Maxwell construction. 
We note that, just as in equilibrium theories for coexistence, the quantitative accuracy of any theory for nonequilibrium coexistence will of course depend on the quality of the equations-of-state, a potential source of the discrepancy in 3d. 

\section*{Nonequilibrium Interfacial Phenomena}

At this point, let us now consider physically why our nonequilibrium mechanical theory consistently predicts a wider binodal when compared to the equilibrium Maxwell construction in the $p-\upsilon$ plane. 
We first note Eq.~(\ref{eq:equilcriteria2}) has a clear mechanical interpretation. 
The integrand $p(\upsilon) - p^{\rm coexist}$ isolates the contribution to the pressure arising solely due to interfacial forces.
The integral can thus be interpreted as the mechanical work exerted by the interfacial forces on a particle as it moves from one phase to the other. 
In equilibrium, this (reversible) work is identically zero: moving a particle from liquid to gas (or gas to liquid) requires no work.
In the case of ABPs, performing the \textit{equilibrium} Maxwell construction in the $p-\upsilon$ plane [with the coexistence pressure $\mathcal{P}^{\rm coexist}$ determined from the nonequilibrium theory, see Fig.~\ref{fig:figure2}(a)] -- the interface \textit{works against} particle removal from the liquid phase:
\begin{equation}
    \label{eq:ABPinsertionwork}
    \mathcal{W}_{\rm interf}^{\rm liq \rightarrow gas} = \int_{\upsilon^{\rm liq}}^{\upsilon^{\rm gas}} \left [\mathcal{P}(\upsilon) - \mathcal{P}^{\rm coexist}\right ] \ d\upsilon \geq 0 \ ,
\end{equation}
where the equality only holds only at the critical point.
This physical picture is consistent with the unique interfacial structure of MIPS, where ABPs within the interface are polarized facing into the liquid phase. 
As activity increases, this interfacial polarization intensifies and so too does the departure from the equilibrium Maxwell construction. 

\begin{figure}[t]
	\centering
	\includegraphics[width=.475\textwidth]{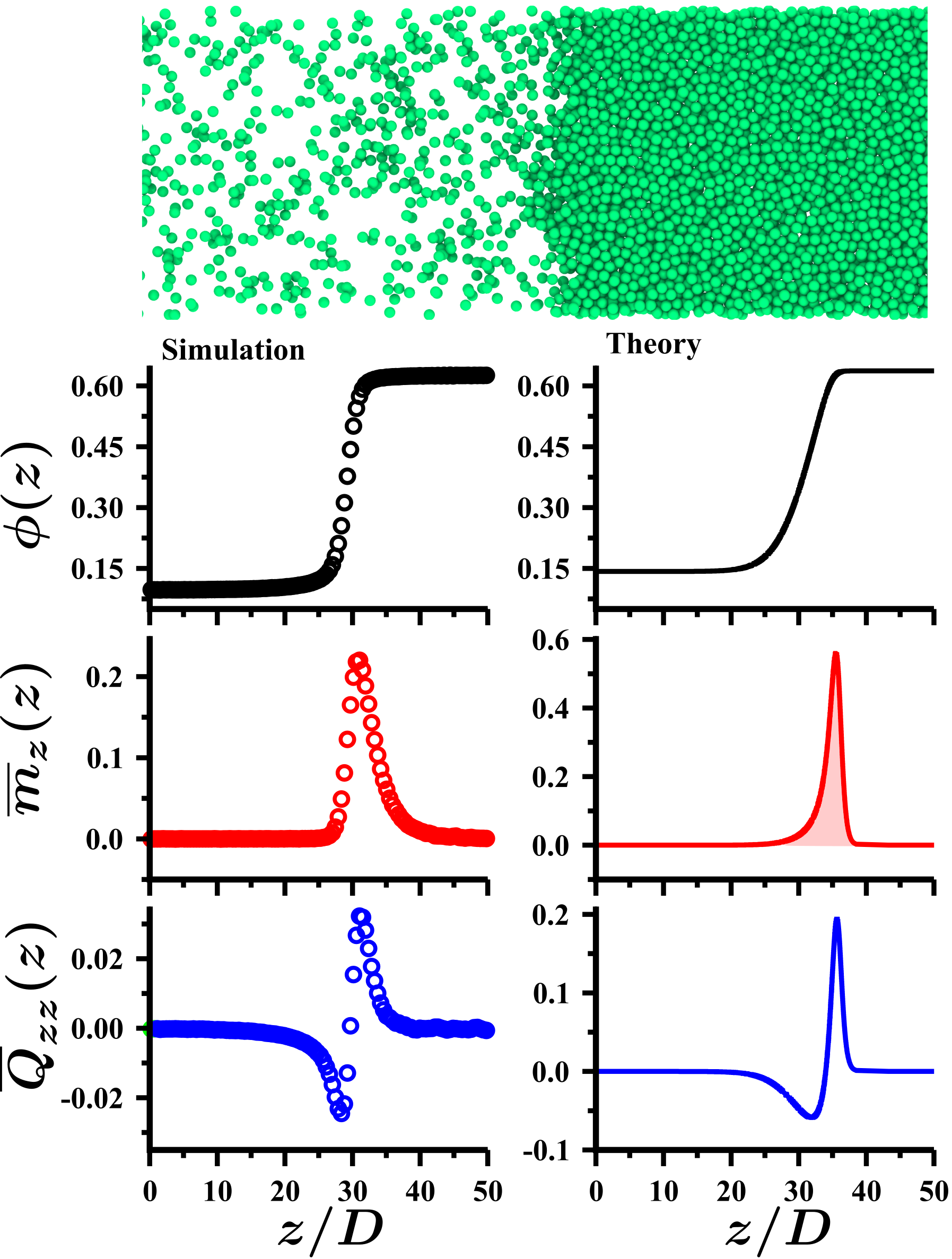}
	\caption{\protect\small{{Comparison of the one-body orientational moments obtained from simulation and theory for 3d ABPs with $\ell_0/D \approx 44.5$. Snapshot represents an instantaneous system configuration. Only a narrow slice (in the out-of-plane direction) of particles are shown for clarity. Polar and nematic order profiles are made dimensionless by the particle volume. Spatial integral of $m_z$ (shaded) is directly proportional to the difference in liquid and gas phase pressures, coupling the interfacial structure to the bulk phase behavior. }}}
	\label{fig:figure4}
\end{figure}

The above discussion makes clear that nonequilibrium interfacial forces play a determining role in the phase behavior of driven systems.
We can investigate this interfacial structure in greater detail as our mechanical theory, by its very nature, makes predictions about the structure of the interface that can be compared with simulation.
{\color{black}
We emphasize that, just as is the case for equilibrium systems, a small gradient theory may fail to \textit{quantitatively} capture the precise structure of the interface while accurately describing the binodal.
A solution of Eq.~(\ref{eq:mechanicaldiffeq}) is shown in Fig.~(\ref{fig:figure4}), where we find good \textit{qualitative} agreement between our mechanical theory and simulation results for the density $\phi$, polarization $m_z$, and nematic order $Q_{zz}$ profiles. 
Additionally, we observe the polar order is proportional to $d\phi/dz$, and the nematic order is proportional to $dm_z/dz$, as predicted by their conservation equations.}

The polarization density, implicated above in the violation of the equilibrium Maxwell construction, can be understood as follows.
From the momentum balance, the difference in $p_{\rm C}$ between the two phases is balanced by the integral of the active force density: $p_{\rm C}(\rho^{\rm liq}) -  p_{\rm C}(\rho^{\rm gas}) = \int_{z^{\rm gas}}^{z^{\rm liq}} \zeta U_0 m_z dz$. 
Particles at the interface are oriented and exert active forces \textit{towards} the phase with a higher interaction pressures or density, suppressing the removal of particles from the liquid phase. 
In the absence of these interfacial active forces (and in the absence of attractive cohesive forces keeping the liquid intact), there would be nothing to prevent the complete dissolution of the liquid phase. 

\begin{figure}[t]
	\centering
	\includegraphics[width=.475\textwidth]{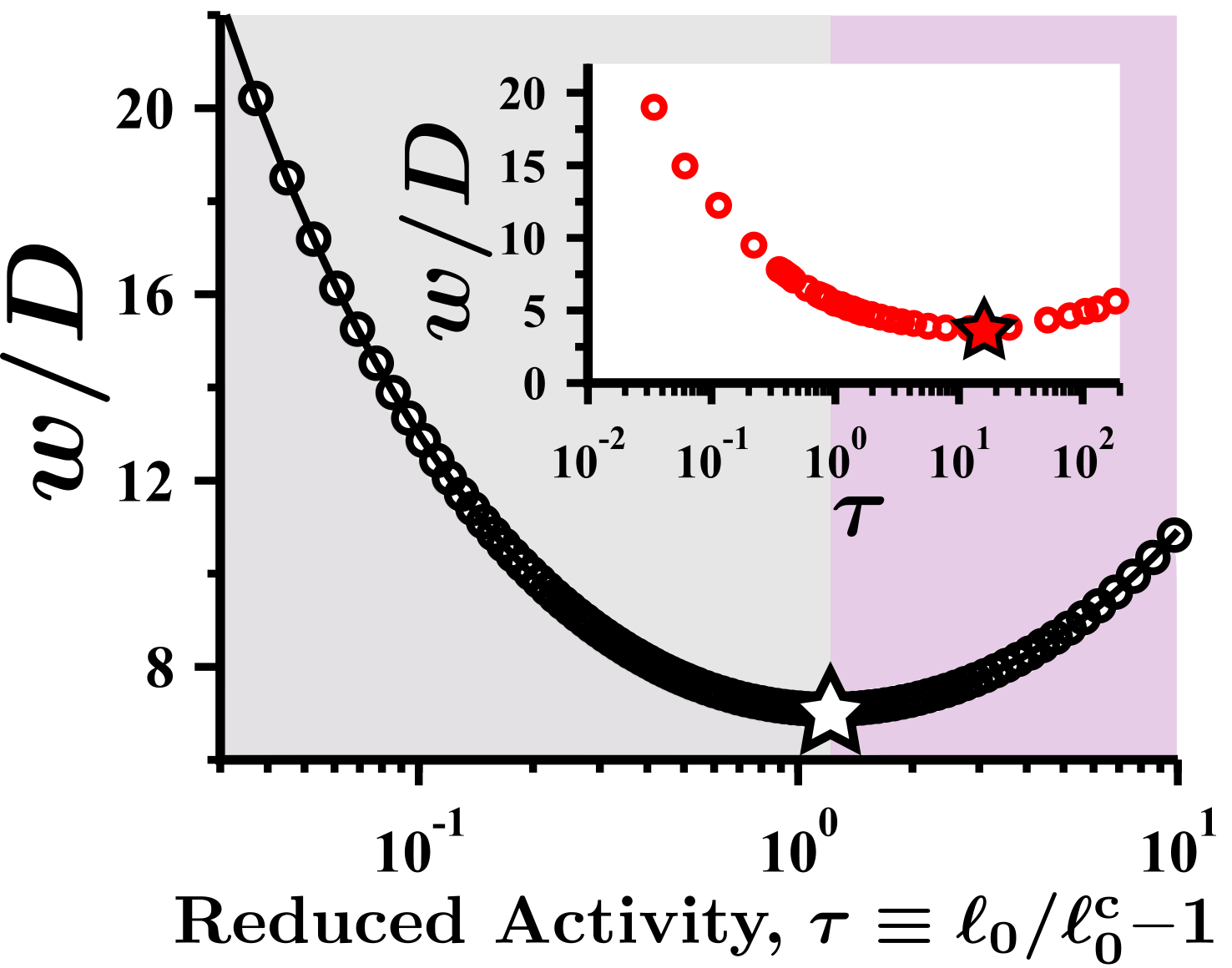}
	\caption{\protect\small{{Theoretical interfacial width $w$ of 3d ABPs as a function of the critical parameter (where $\ell_0^{\rm c}$ is the critical activity) with simulations (inset) corroborating the predicted nonmonotonicity. Stars denote local minima.}}}
	\label{fig:figure5}
\end{figure}

The internally-generated active force density engenders a unique non-monotonic trend in the interfacial width (see Materials and Methods), predicted by our theory (see Fig.~\ref{fig:figure5}).
This behavior was first observed in the simulations of Lauersdorf et al.~\cite{Lauersdorf2021} and reproduced here in our simulations of active spheres (Fig.~\ref{fig:figure5} inset).  
This trend is in stark contrast to interfaces in equilibrium systems where the width of the interface decreases monotonically as the system is taken deeper into the coexistence region.
{\color{black}Again, while a small gradient theory is not expected to quantitatively capture the structure of the interface, our theory is able to capture this effect qualitatively.} 

To illustrate that the origins of this unique nonequilibrium effect are again rooted in the interfacial active force density, consider the following.
As one moves deeper into the two-phase region, the difference in interaction pressures (or densities) between coexisting phases increases, and so must the total active force provided by the particles at the interface to maintain this density difference.  
For sufficiently low activities, the active force required can be achieved by amplifying the active force density, $\zeta U_0 m_z = \zeta U_0 \rho \langle q_z \rangle$, by better alignment of particle orientations $\langle q_z \rangle$ towards the liquid phase, which results in a more compact and thinner interface.
However, this reinforcement mode is limited due to the upper bound of the magnitude of the active force density imposed by perfect alignment $\langle q_z \rangle = 1$. 
To supply the large required active force needed at high activity, the width of the interface must increase with activity -- once a packed layer of particles is fully aligned, more layers are necessary to produce the required active force.

\section*{Discussion and Conclusions}
The nonequilibrium mechanical theory presented in this work allows for the determination of phase diagrams from bulk equations-of-state without making any assumptions regarding the distribution of microstates. 
Our theory identifies the effective pressure $\mathcal {P}$, which includes the pressure arising from conservative interactions and those arising from nonequilibrium body forces, as the critical mechanical quantity in determining the phase behavior of nonequilibrium systems,. 
Using MIPS as a case study, we find that using a true nonequilibrium coexistence theory results in significantly better predictions than the binodal obtained through the naive use of the equilibrium coexistence criteria. 

In equilibrium, the coexistence criteria for phase separation are independent of the system details. 
All that is required is the equation-of-state (the pressure or chemical potential) to determine the phase diagram. 
{\color{black}For nonequilibrium systems, the interfacial stresses must be determined to derive the coexistence criteria, which will generally result in system-specific coexistence criteria [i.e., a system specific $\mathcal{E}(\rho)$]. 
Moreover, while the order at which the density-gradient expansion is truncated for equilibrium systems will not affect $\mathcal{E}(\rho)$, there is no such guarantee for nonequilibrium systems. 
This is a result of the coefficients for a nonequilibrium system generally not emerging from a variational principle as in equilibrium.}
These considerations might suggest that the equilibrium coexistence criteria, while both rigorously and quantitatively incorrect, might at least provide a rough pragmatic estimate for the binodal of a nonequilibrium material~\cite{Takatori2015, Zhang2021}.
However, any departure from the equilibrium Maxwell construction likely indicates the significance of nonequilibrium interfacial forces. 
Indeed, our theory reveals that the internally generated active force density -- present only within the interface -- dictates the interface's structure and, in turn, the appropriate coexistence criteria.

Finally, the mechanical theory for nonequilibrium phase separation presented in this work applies to scenarios where density is the sole order parameter. 
A myriad of other nonequilibrium phase transitions have been observed in recent years, including symmetry-breaking transitions [such as active crystallization~\cite{Omar2021}], transitions with non-conserved order parameters~\cite{Fruchart2020}, and transitions with multiple order parameters, including traveling states{\color{black}~\cite{You2020, Saha2020, Dinelli2022, Chiu2022}.}
A general mechanical theory, such as that developed here, for these and other phase transitions would provide a much-needed framework for constructing and characterizing nonequilibrium coexistence.  
 
\matmethods{
Here, we briefly summarize the simulation and theoretical details while a detailed derivation of the ABP conservation equations is provided in the SI.
\subsection*{Simulations} Particle-based simulations were conducted to determine the binodal for 2d polydisperse disks [equations-of-state for this system were exhaustively determined in Ref.~\cite{Mallory2021}] and the equations-of-state for monodisperse 3d hard spheres [the binodal of this system was determined in Ref.~\cite{Omar2021}].
In all simulations, particles follow the equations-of-motion provided in the main text [Eqs.~(\ref{eq:eqn_motion_trans})~and~(\ref{eq:eqn_motion_rot})] and the interparticle force $\mathbf{F^C}[\mathbf{r}^N; \varepsilon, \sigma]$ is taken to result from a Weeks-Chandler-Anderson (WCA) potential~\cite{Weeks1971} (characterized by a Lennard-Jones diameter $\sigma_{\rm LJ}$ and energy scale $\varepsilon$). 
Despite the use of a continuous potential, hard-particle statistics can be effectively achieved through careful consideration of the different force scales, as discussed in Ref.~\cite{Omar2021}.
Lacking translational Brownian motion, which simply attenuates the influence of activity on the phase behavior, these particles strictly exclude volume with a diameter $D$ set by the potential stiffness $\mathcal{S} \equiv \varepsilon/(\zeta U_0\sigma_{\rm LJ})$ as a measure of the relative strength of conservative and active forces.
Continuous repulsions act only at distances between $D$ and $2^{1/6} \sigma_{\rm LJ}$, a range that quickly becomes negligible as the stiffness $\mathcal{S}$ increases.
We use a stiffness $\mathcal{S}=50$ for which $D/(2^{1/6}\sigma_{\rm LJ}) = 0.9997$, effectively achieving hard-sphere statistics.
We therefore take the diameter to simply be $D = 2^{1/6}\sigma_{\rm LJ}$.
Holding $\mathcal{S}$ fixed to remain in this hard-sphere limit, the system state is independent of the active force magnitude and is fully described by two geometric parameters: the volume fraction $\phi = N\pi D^3/6V$ (or area fraction $\phi = N\pi D^2/4A$) and the dimensionless intrinsic run length $\ell_0 / D$.

All simulations were conducted with a minimum of 54,000 particles using the GPU-enabled \texttt{HOOMD-blue} software package~\cite{Anderson2020}. 
Additional details for the construction of the 3d equations-of-state are provided in the SI.

\subsection*{Fokker-Planck Equation}
The Fokker-Planck (or Smoluchowski) describing the $N$-body distribution of particle positions and orientations has the following form:
\begin{subequations}
\label{eq:fokker_planck}
\begin{equation}
\label{eq:ABP_smoluchowski}
    \frac{\partial f_N}{\partial t} + \sum_\alpha \boldsymbol{\nabla}_\alpha \cdot \mathbf{j}_\alpha^T + \sum_\alpha \boldsymbol{\nabla}_\alpha^R \cdot \mathbf{j}_\alpha^R = 0\ .
\end{equation}
Here, $f_N(\boldsymbol{\Gamma}, t)$ is the probability density of observing a configuration $\boldsymbol{\Gamma} \equiv (\mathbf{r}_1, \mathbf{r}_2,..., \mathbf{r}_N, \mathbf{q}_1, \mathbf{q}_2,..., \mathbf{q}_N)$ at time $t$,
$\mathbf{r}_\alpha$ and $\mathbf{q}_\alpha$ ($|\mathbf{q}_\alpha| = 1$) are the position and orientation vectors of particle $\alpha$, $\mathbf{j}_\alpha^T$ and $\mathbf{j}_\alpha^R$ are translational and rotational fluxes of particle $\alpha$, and $\boldsymbol{\nabla}_\alpha = \partial / \partial \mathbf{r}_\alpha$ and $\boldsymbol{\nabla}_\alpha^R = \mathbf{q}_\alpha \times \partial / \partial \mathbf{q}_\alpha$ are translational and rotational gradient operators. 
 The fluxes are given by
 \begin{equation}
     \mathbf{j}_\alpha^T = U_0 \mathbf{q}_\alpha f_N + \frac{1}{\zeta} \mathbf{F}_\alpha^{\rm C} f_N \ , \label{eq:ABP_j_T}
 \end{equation}
 \begin{equation}
     \mathbf{j}_\alpha^R = -\tau_R^{-1} \boldsymbol{\nabla}_\alpha^R f_N \label{eq:ABP_j_R} \ .
 \end{equation}
 \end{subequations}
The application of our nonequilibrium coexistence theory requires the steady-state (and density flux-free) linear momentum balance and the conservation equations of any field variable appearing in the momentum balance. 
Equation~(\ref{eq:fokker_planck}) and the microscopic definition of the field variables can be used to obtain these conservation equations (see SI for details), which are summarized next. 
\subsection*{Conservation Equations}
Conservation of number density is simply the continuity equation:
\begin{equation}
\label{eq:n_eq_ABP_tensorial}
    \frac{\partial \rho}{\partial t} + \boldsymbol{\nabla} \cdot \mathbf{j}^{\rho} = 0 \ , 
\end{equation}
which is coupled to linear momentum conservation:
\begin{equation}
\label{eq:n_flux_ABP_tensorial}
    \mathbf{0} = \boldsymbol{\nabla} \cdot \boldsymbol{\sigma}^{\rm C}+ \zeta U_0 \mathbf{m} -\zeta \mathbf{j}^{\rho}  \ .
\end{equation}
The polar order field $\mathbf{m}(\mathbf{x}, t)$ satisfies its own conservation equation: 
\begin{subequations}
\begin{equation}
\label{eq:m_eq_ABP_tensorial}
    \frac{\partial \mathbf{m}}{\partial t} + \boldsymbol{\nabla} \cdot \mathbf{j^m} + \frac{d-1}{\tau_R} \mathbf{m}  = \mathbf{0} \ , 
\end{equation}
where the polarization flux follows:
\begin{equation}
\label{eq:m_flux_ABP_tensorial}
    \mathbf{j^m} = U_0 \overline{U} \left(\mathbf{Q} + \frac{1}{d} \rho \mathbf{I} \right)\ .
\end{equation}
\end{subequations}
A microscopic expression for the dimensionless average active speed $\overline{U}$ is provided in the SI. 
An additional term, not included in Eq.~(\ref{eq:m_flux_ABP_tensorial}), also appears but is found to have only a negligible quantitative effect on our findings as detailed in the SI. 

The nematic order conservation and constitutive equations are found to be:
\begin{subequations}
\begin{equation}
\label{eq:Q_eq_ABP_tensorial}
    \frac{\partial \mathbf{Q}}{\partial t} + \boldsymbol{\nabla} \cdot \mathbf{j^Q} + \frac{2d}{\tau_R} \mathbf{Q} = \mathbf{0} \ , 
\end{equation}
\begin{equation}
\label{eq:Q_flux_ABP_tensorial}
    \mathbf{j^Q} = U_0 \overline{U} \mathbf{B} + U_0 \mathbf{m} \cdot \left(\frac{\overline{U}}{d + 2}  \boldsymbol{\alpha}  - \frac{1}{d} \mathbf{I}\mathbf{I} \right)  - \frac{1}{d\zeta} \boldsymbol{\nabla} \cdot \boldsymbol{\sigma}^{\rm C} \mathbf{I} \ , 
\end{equation}
\end{subequations}
where $\boldsymbol{\alpha}$ is an isotropic fourth-rank tensor. 
(In indicial notation, $\alpha_{ijkl} = \delta_{ij}\delta_{kl} + \delta_{ik}\delta_{jl} + \delta_{il}\delta_{jk}$ where $\delta_{ij}$ is the second-rank identity tensor.)
In Eq.~(\ref{eq:Q_flux_ABP_tensorial}), the microscopic expression for $\overline{U}$ differs from that in Eq.~(\ref{eq:m_flux_ABP_tensorial}). 
However, to good approximation, these speeds can be taken to be the same, allowing us to express the steady-state equations with only two equations-of-state: $p_{\rm C}$ and $\overline{U}$.
{\color{black}
\subsection*{Interfacial Width Definition}
The interfacial width is not a uniquely defined quantity. 
Here, for both our theory and simulations, we compute the interfacial width using the ensemble-averaged density profile, $\phi(z)$. 
We seek a definition of interfacial width which does not presume a particular functional form of $\phi(z)$.
We therefore use the ``10-90 thickness''~\cite{Lekner1978} definition of interfacial width, which defines the width as the distance between the two locations, $z_1$ and $z_2$ (i.e., w = $|z_2-z_1|$), at which  $\phi(z_1) = \phi^{\rm gas} + 0.1(\phi^{\rm liq} - \phi^{\rm gas})$ and $\phi(z_2) = \phi^{\rm gas} + 0.9(\phi^{\rm liq} - \phi^{\rm gas})$. 
The qualitative results were found to be insensitive to the precise definition of interfacial width.
}
}

\showmatmethods{}

\acknow{
A.K.O. is deeply indebted to Phill Geissler for his numerous insights regarding this work.
We thank Katie Klymko, Karol Makuch, Yizhi Shen, Zhiwei Peng, Andy Ylitalo, Dan Evans, and Luke Langford for helpful discussions.
We gratefully acknowledge support from the Schmidt Science Fellowship in partnership with the Rhodes Trust (A.K.O.), Kwanjeong Educational Foundation (H.R.), Arnold and Mabel Beckman Foundation (S.A.M.), and National Science Foundation under Grant No. CBET-1803662 (J.F.B.).
}
\showacknow{}

\end{document}


\title{Supporting Information: Mechanical Theory of Nonequilibrium Coexistence and Motility-Induced Phase Separation}

\author{Ahmad K. Omar}
\email{aomar@berkeley.edu}
\affiliation{Department of Materials Science and Engineering, University of California, Berkeley, California 94720, USA}
\affiliation{Materials Sciences Division, Lawrence Berkeley National Laboratory, Berkeley, California 94720, USA}
\author{Hyeongjoo Row}
\affiliation{Division of Chemistry and Chemical Engineering, California Institute of Technology, Pasadena, California 91125, USA}
\author{Stewart A. Mallory}
\affiliation{Department of Chemistry, The Pennsylvania State University, University Park, Pennsylvania 16802, USA}
\author{John F. Brady}
\email{jfbrady@caltech.edu}
\affiliation{Division of Chemistry and Chemical Engineering, California Institute of Technology, Pasadena, California 91125, USA}

\maketitle

\section{Active Brownian Particle Conservation Equations and Closures}
We provide a derivation of the equations presented in the main text: the conservation equations for density, linear momentum, and the other orientational moment fields for a collection of $N$ interacting active Brownian particles (ABPs).
We emphasize that these equations for interacting ABPs have appeared in various forms throughout the literature~\cite{Paliwal2018, Solon2018, Epstein2019}.
However, in addition to deriving the equations, we will introduce the closures and assumptions that allow us to determine the stationary states of coexistence. 

The Fokker-Planck (or Smoluchowski) equation describing the $N$-body distribution of particle positions and orientations follows from the particle equations-of-motion and is given by:
\begin{subequations}
\label{eq:Fokker-Planck equation}
\begin{equation}
    \frac{\partial f_N(\boldsymbol{\Gamma}, t)}{\partial t} + \sum_\alpha \boldsymbol{\nabla}_\alpha \cdot \mathbf{j}_\alpha^T + \sum_\alpha \boldsymbol{\nabla}_\alpha^R \cdot \mathbf{j}_\alpha^R = 0\ .  \label{eq:original_smolu}
\end{equation}
Here, $f_N(\boldsymbol{\Gamma}, t)$ is the probability density of observing a configuration $\boldsymbol{\Gamma} \equiv (\mathbf{r}_1,\ldots, \mathbf{r}_N, \mathbf{q}_1,\ldots, \mathbf{q}_N)$ at time $t$, $\mathbf{r}_\alpha$ and $\mathbf{q}_\alpha$ $\left(|\mathbf{q}_\alpha| = 1\right)$ are the position and orientation vectors of particle $\alpha$, $\mathbf{j}_\alpha^T$ and $\mathbf{j}_\alpha^R$ are the translational and rotational fluxes of particle $\alpha$, and $\boldsymbol{\nabla}_\alpha = \partial / \partial \mathbf{r}_\alpha$ and $\boldsymbol{\nabla}_\alpha^R = \mathbf{q}_\alpha \times \partial / \partial \mathbf{q}_\alpha$ are translational and rotational gradient operators. 
The fluxes are given by:
 \begin{equation}
     \mathbf{j}_\alpha^T = U_0 \mathbf{q}_\alpha f_N + \frac{1}{\zeta} \mathbf{F}_\alpha f_N - D_T 
     \boldsymbol{\nabla}_{\alpha} f_N \ , \label{eq:original_j_T}
 \end{equation}
 \begin{equation}
     \mathbf{j}_\alpha^R = -\frac{1}{\tau_R} \boldsymbol{\nabla}_\alpha^R f_N, \label{eq:original_j_R} \ ,
 \end{equation}
 \end{subequations}
where $U_0$ is the intrinsic active speed, $\zeta$ is the translational drag coefficient, $D_T$ is the translational Brownian diffusivity (neglected in the main text but provided here for completeness), $\tau_R$ is the reorientation time scale, and $\mathbf{F}_\alpha$ is the conservative force on particle $\alpha$ arising from a the potential energy $U(\mathbf{r}_1,\ldots, \mathbf{r}_N)$. 
We consider pairwise additive isotropic interparticle interactions such that $U(\mathbf{r}_1,\ldots, \mathbf{r}_N) = \sum_\alpha U^{\rm ext}(\mathbf{r}_\alpha) + \sum_\alpha \sum_{\beta \neq \alpha} U_2(r_{\alpha\beta}) / 2$. 
Here, $U^{\rm ext}(\mathbf{x})$ is the externally imposed potential at position $\mathbf{x}$, $U_2(r)$ is the pair interaction potential, and $\mathbf{r}_{\alpha \beta} = \mathbf{r}_\alpha - \mathbf{r}_\beta$ ($r_{\alpha \beta} = |\mathbf{r}_{\alpha \beta}|$). 
The force on particle $\alpha$ arising from the conservative potentials thus has two contributions $\mathbf{F}_{\alpha} = \mathbf{F}_{\alpha}^{\rm ext} + \mathbf{F}_{\alpha}^{\rm C}$, where $\mathbf{F}_{\alpha}^{\rm ext} = -\boldsymbol{\nabla}_{\alpha} U^{\rm ext}(\mathbf{r}_\alpha)$, $\mathbf{F}_{\alpha}^{\rm C} = \sum_{\beta \neq \alpha} \mathbf{F}_{\alpha\beta}^{\rm C}$, and $\mathbf{F}_{\alpha\beta}^{\rm C} = - \boldsymbol{\nabla}_{\alpha} U_2(r_{\alpha \beta})$. 
From Eq.~\eqref{eq:Fokker-Planck equation} the dynamical operator is defined as $\mathcal{L} \equiv \sum_\alpha [ \boldsymbol{\nabla}_{\alpha}\cdot(-U_0 \mathbf{q}_\alpha - \zeta^{-1} \mathbf{F}_\alpha + D_T \boldsymbol{\nabla}_{\alpha}) + \boldsymbol{\nabla}_\alpha^R \cdot(\tau_R^{-1} \boldsymbol{\nabla}_\alpha^R)]$ such that $\partial f_N / \partial t = \mathcal{L} f_N$.

The full $N$-body distribution function $f_N$ allows for the determination of the ensemble average of any observable $\mathcal{O}$: $\mathcal{O} = \langle\hat{\mathcal{O}}(\boldsymbol{\Gamma}) \rangle$, where $\langle\ \cdot\ \rangle \equiv \int (\ \cdot\ ) f_N(\boldsymbol{\Gamma}, t) \ d\boldsymbol{\Gamma}$ is the ensemble average and $\hat{\mathcal{O}}(\boldsymbol{\Gamma})$ is the microscopic definition of the observable $\mathcal{O}$. 
The time evolution of an observable is then:
\begin{equation}
    \frac{\partial \mathcal{O}}{\partial t} = \int \hat{\mathcal{O}} \left(\frac{\partial f_N}{\partial t}\right) \ d\boldsymbol{\Gamma} = \int \hat{\mathcal{O}} (\mathcal{L}f_N) \ d\boldsymbol{\Gamma} = \int (\mathcal{L}^{\rm \star} \hat{\mathcal{O}}) f_N \ d\boldsymbol{\Gamma} = \langle \mathcal{L}^{\rm \star} \hat{\mathcal{O}} \rangle\ , \label{eq:adjoint_propagation} 
\end{equation}
where, in the case of Eq.~\eqref{eq:Fokker-Planck equation}, the adjoint of $\mathcal{L}$ is $\mathcal{L}^{\rm \star} \equiv \sum_\alpha [ (U_0 \mathbf{q}_\alpha + \zeta^{-1} \mathbf{F}_\alpha + D_T \boldsymbol{\nabla}_{\alpha}) \cdot \boldsymbol{\nabla}_{\alpha}  + \tau_R^{-1} \boldsymbol{\nabla}_\alpha^R \cdot \boldsymbol{\nabla}_\alpha^R]$. 

We first derive an evolution equation of the (number) density field $\rho(\mathbf{x}, t) = \langle\hat{\rho}(\mathbf{x})\rangle$, where $\hat{\rho}(\mathbf{x}) = \sum_\alpha \delta(\mathbf{x} - \mathbf{x}_\alpha)$ is the microscopic density of particles. 
The continuity equation directly follows from this procedure:
\begin{subequations}
\label{eq:rho_evolution}
\begin{equation}
    \frac{\partial \rho}{\partial t} + \boldsymbol{\nabla} \cdot \mathbf{j}^\rho = 0\ , \label{eq:rho_eqn}
\end{equation}
where:
\begin{equation}
    \mathbf{j}^\rho = U_0 \mathbf{m}(\mathbf{x}, t) +   \frac{1}{\zeta} \mathbf{F}^{\rm ext}(\mathbf{x}) \rho +\frac{1}{\zeta} \boldsymbol{\nabla} \cdot \boldsymbol{\sigma}^{\rm C}(\mathbf{x}, t) - D_T \boldsymbol{\nabla} \rho\ . \label{eq:rho_flux}
\end{equation}
\end{subequations}
Here, $\boldsymbol{\nabla} \equiv \partial / \partial \mathbf{x}$ is the spatial gradient operator, $\mathbf{j}^\rho$ is the particle flux,  $\mathbf{m}(\mathbf{x}, t) \equiv \langle \hat{\mathbf{m}}(\mathbf{x})\rangle$ is the polarization density field, $\hat{\mathbf{m}}(\mathbf{x}) = \sum_\alpha \mathbf{q}_\alpha \delta (\mathbf{x} - \mathbf{r}_\alpha)$ is the microscopic density of polarization, and:
\begin{equation}
    \boldsymbol{\sigma}^{\rm C}(\mathbf{x}, t) = \left\langle -\frac{1}{2} \sum_\alpha \sum_{\alpha \neq \beta} \mathbf{r}_{\alpha\beta} \mathbf{F}_{\alpha \beta}^{\rm C} b_{\alpha \beta} \right\rangle \ , \label{eq:stress}
\end{equation}
is the stress generated by the pairwise interparticle forces and $b_{\alpha \beta}(\mathbf{x}; \mathbf{r}_{\alpha}, \mathbf{r}_{\beta})  = \int_{0}^{1} \delta(\mathbf{x} -  \mathbf{r}_{\beta} - \lambda \mathbf{r}_{\alpha \beta}) \ d\lambda$ is the bond function~\cite{Noll1955, Lehoucq2010}.

The four terms in the particle flux~(\ref{eq:rho_flux}) correspond to the four modes of particle transport: transport driven by the active force, external forcing, interparticle forces, and Brownian motion.
Equation~\eqref{eq:rho_flux} is also a statement of linear momentum conservation. 
To see this we rearrange and find:
\begin{subequations}
\label{eq:linmomentumdefs}
\begin{equation}
   \mathbf{0} = \boldsymbol{\nabla} \cdot \boldsymbol{\sigma}(\mathbf{x}, t) + \mathbf{b}(\mathbf{x}, t) \ ,
   \label{eq:overdamped_momentum_genera}
\end{equation}
where the body forces are the terms in Eq.~\ref{eq:rho_flux} that \textit{generally} may not be expressed as divergences of a tensor: 
\begin{equation}
   \mathbf{b}(\mathbf{x}, t) =   -\zeta \mathbf{j}^\rho(\mathbf{x}, t) + \zeta U_0 \mathbf{m}(\mathbf{x}, t) + \mathbf{F}^{\rm ext}(\mathbf{x}) \rho(\mathbf{x}, t) \ ,  
   \label{eq:body_def}
\end{equation}
and the stresses are:
\begin{equation}
   \boldsymbol{\sigma}(\mathbf{x}, t) =  \boldsymbol{\sigma}^{\rm C}(\mathbf{x}, t) - \zeta D_T \rho(\mathbf{x}, t) \mathbf{I}\ ,  
   \label{eq:stress_def}
\end{equation}
where $-\zeta D_T \rho\mathbf{I}$ represents the Brownian ``ideal gas'' stress  and $\mathbf{I}$ is the second-rank identity tensor.
\end{subequations}

In order to solve Eq.~\eqref{eq:rho_evolution}, we require the polarization density $\mathbf{m}$ and the stress $\boldsymbol{\sigma}^{\rm C}$. 
The polarization density has its own evolution equation. 
By letting $\hat{\mathcal{O}} = \hat{\mathbf{m}}$ in Eq.~\eqref{eq:adjoint_propagation} and following procedures similar to what were applied in order to obtain Eq.~\eqref{eq:rho_evolution}, we find:
\begin{subequations}
\label{eq:polarization_evolution}
\begin{equation}
    \frac{\partial \mathbf{m}}{\partial t} + \boldsymbol{\nabla} \cdot \mathbf{j}^\mathbf{m} + \frac{d-1}{\tau_R}\mathbf{m}  = 0\ , \label{eq:m_eqn}
\end{equation}
\begin{equation}
    \mathbf{j}^\mathbf{m} = U_0 \mathbf{\tilde{Q}}(\mathbf{x}, t) +  \frac{1}{\zeta} \mathbf{F}^{\rm ext}\mathbf{m}
    + \frac{1}{\zeta} \boldsymbol{\kappa}^{\mathbf{m}} (\mathbf{x}, t)
    + \frac{1}{\zeta} \boldsymbol{\nabla} \cdot \boldsymbol{\Sigma}^{\mathbf{m}} (\mathbf{x}, t) - D_T \boldsymbol{\nabla} \mathbf{m} \ , \label{eq:m_flux}
\end{equation}
\end{subequations}
where $\mathbf{j}^{\mathbf{m}}$ is the polarization flux, $d$ is the spatial dimension, and $\mathbf{\tilde{Q}}(\mathbf{x}, t) \equiv \langle \mathbf{\hat{Q}}(\mathbf{x})\rangle$ is the nematic order density field (where $\mathbf{\hat{Q}}(\mathbf{x}) = \sum_\alpha \mathbf{q}_\alpha\mathbf{q}_\alpha \delta (\mathbf{x} - \mathbf{r}_\alpha)$ is the microscopic nematic density). 
The interparticle forces contribute to the transport of the polarization in body-force-like (i.e.,~no divergence) $\boldsymbol{\kappa}^{\mathbf{m}}$ and stress-like $\boldsymbol{\Sigma}^{\mathbf{m}}$ manners, which are defined as:
\begin{equation}
    \boldsymbol{\kappa}^{\mathbf{m}}(\mathbf{x}, t) 
    = \left\langle 
    \frac{1}{2} \sum_\alpha \sum_{\alpha \neq \beta}  \mathbf{F}_{\alpha \beta}^{\rm C} (\mathbf{q}_{\alpha} - \mathbf{q}_{\beta})\delta(\mathbf{x}-\mathbf{r}_\alpha)
    \right\rangle \ , \label{eq:kappa_m}
\end{equation}
\begin{equation}
    \boldsymbol{\Sigma}^{\mathbf{m}}(\mathbf{x}, t) = 
    \left\langle 
    -\frac{1}{2} \sum_\alpha \sum_{\alpha \neq \beta}  \mathbf{r}_{\alpha\beta} \mathbf{F}_{\alpha \beta}^{\rm C} \mathbf{q}_{\alpha} b_{\alpha \beta}
    \right\rangle \ . \label{eq:Sig_m}
\end{equation}
Equation~\eqref{eq:Sig_m} makes clear that $d\mathbf{S}(\mathbf{x}) \cdot \boldsymbol{\Sigma}^{\mathbf{m}}/\zeta$ is the average transport of polarization due to the interparticle forces acting \textit{across} the infinitesimal area $dS$ from the direction of $d\mathbf{S}$. 
From the definition [Eq.~\eqref{eq:kappa_m}] of the body-force-like term $\boldsymbol{\kappa}^{\mathbf{m}}$, we observe that configurations with $\mathbf{q}_\alpha\!=\!\mathbf{q}_\beta$ do not contribute to $\boldsymbol{\kappa}^{\mathbf{m}}$ and configurations with $\mathbf{q}_\alpha\!=\!-\mathbf{q}_\beta$ contribute the most to $\boldsymbol{\kappa}^{\mathbf{m}}$ in magnitude. 
This observation is indicative that $\boldsymbol{\kappa}^{\mathbf{m}}$ is correlated with the reduction in the effective active speed $U_{\rm eff}$ due to interparticle interactions --- a pair of particles slow down when they collide head to head but active motion is largely unaffected when interacting particles are oriented in the same direction. 
From the scaling analysis of the active pressure $p_{\rm act} \sim \rho \zeta U_0 \tau_{R} U_{\rm eff}$~\cite{Takatori2014} and recognizing that the active pressure is proportional to the trace of the polarization flux (with $\mathbf{j}^\mathbf{m} \sim \rho \left\langle \dot{\mathbf{x}} \mathbf{q}\right\rangle$) such that $p_{\rm act} \sim \rho \zeta U_0 \tau_{R} \left\langle \dot{\mathbf{x}} \cdot \mathbf{q}\right\rangle $~\cite{, Patch2018, Solon2018, Das2019, Omar2020, Mallory2021}, we can indeed identify that $\boldsymbol{\kappa}^{\mathbf{m}}$ is directly related to the reduction in the effective speed of active transport of polarization. 
This motivates a constitutive equation 
\begin{equation}
\boldsymbol{\kappa}^{\mathbf{m}} = -\zeta(U_0 - U_{\rm eff}^{\mathbf{m}}) \mathbf{\tilde{Q}}\ , \label{eq:U_bar_m}
\end{equation}
which leads to
\begin{equation}
    \mathbf{j}^\mathbf{m} = U_0 \overline{U}^{\mathbf{m}} \mathbf{\tilde{Q}} +  \frac{1}{\zeta} \mathbf{F}^{\rm ext}\mathbf{m}
    + \frac{1}{\zeta} \boldsymbol{\nabla} \cdot \boldsymbol{\Sigma}^{\mathbf{m}} - D_T \boldsymbol{\nabla} \mathbf{m} \ , \label{eq:m_flux_new}
\end{equation}
where $U_{\rm eff}^{\mathbf{m}} = U_0 \overline{U}^{\mathbf{m}}$ is the effective speed of active polarization transport. 
The dimensionless quantity $\overline{U}^{\mathbf{m}} (\in [0,1])$ represents the effective speed relative to the intrinsic speed $U_0$ and is an equation-of-state depending on the system volume (area) fraction $\phi$ and activity $\ell_0 / D$.
$\overline{U}^{\mathbf{m}} \approx 1$ when particles move nearly freely at low ($\phi \ll 1$) densities and $\overline{U}^{\mathbf{m}} \approx 0$ when particles mobility is limited due to interparticle interactions.

To close our equations, expressions for $\boldsymbol{\sigma}^{\rm C}$, $\mathbf{\tilde{Q}}$, $\overline{U}^{\mathbf{m}}$, and $\boldsymbol{\Sigma}^{\mathbf{m}}$ are required. 
The nematic order density $\mathbf{\tilde{Q}}$ follows its own evolution equation which can again be derived from Eq.~\eqref{eq:adjoint_propagation} with $\hat{\mathcal{O}} = \mathbf{\hat{Q}}$:
\begin{subequations}\label{eq:Qtilde_evolution}
\begin{equation}
    \frac{\partial \mathbf{\tilde{Q}}}{\partial t} + \boldsymbol{\nabla} \cdot \mathbf{j}^{\mathbf{\tilde{Q}}} + \frac{2d}{\tau_R}  \left(\mathbf{\tilde{Q}} - \frac{1}{d} \rho \mathbf{I} \right) = \mathbf{0} \ , \label{eq:Qtilde_eqn}
\end{equation}
\begin{equation}
    \mathbf{j}^{\mathbf{\tilde{Q}}}= U_0 \mathbf{\tilde{B}}(\mathbf{x}, t)
    + \frac{1}{\zeta} \mathbf{F}^{\rm ext}\mathbf{\tilde{Q}}
    + \frac{1}{\zeta} \boldsymbol{\kappa}^{\mathbf{\tilde{Q}}} (\mathbf{x}, t)
    + \frac{1}{\zeta} \boldsymbol{\nabla} \cdot \boldsymbol{\Sigma}^{\mathbf{\tilde{Q}}} (\mathbf{x}, t) 
    - D_T \boldsymbol{\nabla} \mathbf{\tilde{Q}}\ . \label{eq:Qtilde_flux_original}
\end{equation}
\end{subequations}
Here, $\mathbf{j}^{\mathbf{\tilde{Q}}}$ is the nematic order flux and $\mathbf{\tilde{B}}(\mathbf{x}, t) \equiv \langle\sum_\alpha \mathbf{q}_\alpha\mathbf{q}_\alpha\mathbf{q}_\alpha \delta (\mathbf{x} - \mathbf{r}_\alpha)\rangle$. 
Interparticle interactions again result in body-force- and stress-like terms in the nematic order flux:
\begin{equation}
    \boldsymbol{\kappa}^{\mathbf{\tilde{Q}}}(\mathbf{x}, t) 
    = \left\langle 
    \frac{1}{2} \sum_\alpha \sum_{\alpha \neq \beta}  \mathbf{F}_{\alpha \beta}^{\rm C} (\mathbf{q}_{\alpha}\mathbf{q}_{\alpha} - \mathbf{q}_{\beta}\mathbf{q}_{\beta})\delta(\mathbf{x} - \mathbf{r}_\alpha)
    \right\rangle \ , \label{eq:kappa_Q}
\end{equation}
\begin{equation}
    \boldsymbol{\Sigma}^{\mathbf{\tilde{Q}}}(\mathbf{x}, t) = 
    \left\langle 
    -\frac{1}{2} \sum_\alpha \sum_{\alpha \neq \beta}  \mathbf{r}_{\alpha\beta} \mathbf{F}_{\alpha \beta}^{\rm C} \mathbf{q}_{\alpha}\mathbf{q}_{\alpha} b_{\alpha \beta}
    \right\rangle \ . \label{eq:Sig_Q}
\end{equation}
Again, the stress-like term $\boldsymbol{\Sigma}^{\mathbf{\tilde{Q}}}(\mathbf{x}, t)$ represents the average transport of nematic order due to interparticle forces acting across $\mathbf{x}$ and $\boldsymbol{\kappa}^{\mathbf{\tilde{Q}}}$ is related to the reduction in the effective speed of active transport of the nematic order. 
We again propose a constitutive relation:
\begin{equation}
\boldsymbol{\kappa}^{\mathbf{\tilde{Q}}} = -\zeta(U_0 - U_{\rm eff}^{\tilde{Q}}) \mathbf{\tilde{B}}\ . \label{eq:U_bar_Q}
\end{equation}
Consequently, the nematic order flux becomes:
\begin{equation}
    \mathbf{j}^{\mathbf{\tilde{Q}}}= U_0\overline{U}^{\mathbf{\tilde{Q}}} \mathbf{\tilde{B}}(\mathbf{x}, t)
    + \frac{1}{\zeta} \mathbf{F}^{\rm ext}\mathbf{\tilde{Q}}
    + \frac{1}{\zeta} \boldsymbol{\nabla} \cdot \boldsymbol{\Sigma}^{\mathbf{\tilde{Q}}} (\mathbf{x}, t) 
    - D_T \boldsymbol{\nabla} \mathbf{\tilde{Q}}\ , \label{eq:Qtilde_flux_new}
\end{equation}
where $U_{\rm eff}^{\mathbf{\tilde{Q}}} = U_0 \overline{U}^{\mathbf{\tilde{Q}}}$ is the effective speed of active nematic order transport. 
The dimensionless quantity $\overline{U}^{\mathbf{\tilde{Q}}}$ is again an equation-of-state depending on $\phi$ and  $\ell_0 / D$.

It proves convenient to rewrite our equations with the traceless tensorial orientational moments to exclude the portions that are dependent on the lower order orientational moments (i.e.,~$\rho$ and $\mathbf{m}$).
The traceless nematic order is defined as $\mathbf{Q} = \mathbf{\tilde{Q}} - \rho \mathbf{I} / d$ while $\mathbf{B} = \mathbf{\tilde{B}} - \boldsymbol{\alpha} \cdot \mathbf{m} / (d + 2)$, where  $\boldsymbol{\alpha}$ is a fourth-rank isotropic tensor. 
(In indicial notation, $\alpha_{ijkl} = \delta_{ij}\delta_{kl} + \delta_{ik}\delta_{jl} + \delta_{il}\delta_{jk}$ where $\delta_{ij}$ is the second-rank identity tensor.) 
The polarization flux [Eq.~\eqref{eq:m_flux_new}] and nematic order evolution equation [Eqs.~\eqref{eq:Qtilde_eqn} and \eqref{eq:Qtilde_flux_new}] become:
\begin{equation}
    \mathbf{j}^\mathbf{m} = U_0 \overline{U}^{\mathbf{m}} \left(\mathbf{Q} + \frac{1}{d}\rho \mathbf{I}\right) +  \frac{1}{\zeta} \mathbf{F}^{\rm ext}\mathbf{m}
    + \frac{1}{\zeta} \boldsymbol{\nabla} \cdot \boldsymbol{\Sigma}^{\mathbf{m}} - D_T \boldsymbol{\nabla} \mathbf{m} \ , \label{eq:m_flux_new2}
\end{equation}
\begin{subequations}\label{eq:Q_evolution}
\begin{equation}
    \frac{\partial \mathbf{Q}}{\partial t} + \boldsymbol{\nabla} \cdot \mathbf{j}^{\mathbf{Q}} + \frac{2d}{\tau_R} \mathbf{Q}  = \mathbf{0}\ , \label{eq:Q_evolution_eqn}
\end{equation}
where the traceless nematic flux $\mathbf{j}^{\mathbf{Q}} = \mathbf{j}^{\mathbf{\tilde{Q}}} - \frac{1}{d} \mathbf{j}^{\rho}\mathbf{I}$ is:
\begin{equation}
    \mathbf{j}^{\mathbf{Q}} = U_0\overline{U}^{\mathbf{\tilde{Q}}} \left(\mathbf{B} + \frac{1}{d+2}\boldsymbol{\alpha}\cdot\mathbf{m}\right) - \frac{1}{d}U_0\mathbf{m}\mathbf{I} + \frac{1}{\zeta} \mathbf{F}^{\rm ext}\mathbf{Q} + \frac{1}{\zeta} \boldsymbol{\nabla} \cdot \boldsymbol{\Sigma}^{\mathbf{\tilde{Q}}}  - \frac{1}{\zeta d} \boldsymbol{\nabla} \cdot \boldsymbol{\sigma}^{\rm C} \mathbf{I} - D_T \boldsymbol{\nabla} \mathbf{Q}\ . \label{eq:Q_flux}
\end{equation}
\end{subequations}

Now, we have coupled evolution equations for the density [Eq.~\eqref{eq:rho_evolution}], polar order [Eqs.~\eqref{eq:m_eqn} and \eqref{eq:m_flux_new2}], and nematic order [Eq.~\eqref{eq:Q_evolution}]. 
Closing these equations will require expressions for the following unknowns: $\mathbf{B}$, $\boldsymbol{\Sigma}^{\mathbf{\tilde{Q}}}$, $\boldsymbol{\Sigma}^{\mathbf{m}}$, $\boldsymbol{\sigma}^{\rm C}$, $\overline{U}^{\mathbf{m}}$, and $\overline{U}^{\mathbf{\tilde{Q}}}$. 
Importantly, since our mechanical theory for phase coexistence requires density gradients only up to second order, $\mathbf{B}$ and $\boldsymbol{\Sigma}^{\mathbf{\tilde{Q}}}$ can be safely discarded as they will contribute third order ($\mathcal{O}(k^3)$ in the Fourier space) gradient terms. 
We postulate that the effective speeds of active transport for polar and nematic orders are identical, i.e. $\overline{U}(\ell_0/D, \phi) \equiv \overline{U}^{\mathbf{m}} = \overline{U}^{\mathbf{\tilde{Q}}}$. 

The conservative interaction stress $\boldsymbol{\sigma}^{\rm C}$ is a familiar quantity. 
In the absence of spatial gradients, its sole contribution is the isotropic pressure arising from interparticle interactions $\boldsymbol{\sigma}^{\rm C} = -p_{\rm C}(\ell_0/D, \phi) \mathbf{I}$, which again will simply depend on activity and the particle density.
The stress will of course also have gradient terms (the Korteweg stresses).
However, these Korteweg stresses, which arise due to the distortion of the pair-distribution function in the presence of density gradients, are significantly smaller than the gradient terms generated by the active stress.
Indeed, the nonisotropic contribution to the Korteweg stress (which has the same scale as the isotropic contributions, see main text) was measured in Ref.~\cite{Omar2020} in order to compute the surface tension of phase-separated ABPs.
These Korteweg stresses were found to be negligibly small, scaling as $\sim \zeta U_0D$ while the active stress gradient terms scale as $\sim \zeta U_0\ell_0$.
As we are interested in scenarios where MIPS occurs (i.e.,~$\ell_0/D \gg 1$), we neglect the gradient terms arising from the conservative stress such that $\boldsymbol{\sigma}^{\rm C} = -p_{\rm C}(\ell_0/D, \phi) \mathbf{I}$.

\begin{figure}[t]
	\centering
	\includegraphics[width=.75\textwidth]{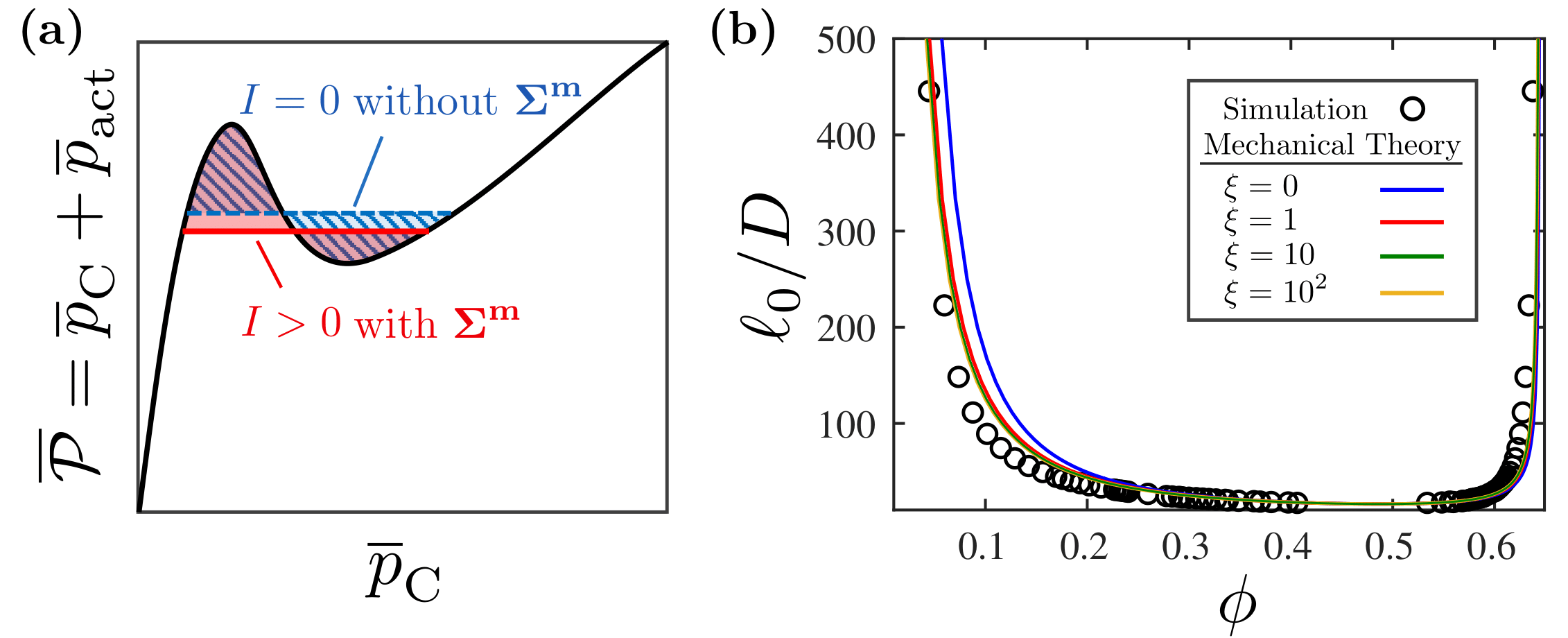}
	\caption{(a) Schematic illustrating the effect of $\boldsymbol{\Sigma}^{\mathbf{m}}$ to the coexistence effective pressure $\mathcal{P}^{\rm coexist}$. On the $\mathcal{P}-p_{\rm C}$ diagram, the integral $I$ [Eq.~\eqref{eq:Integral_epsilon_pc}] represents the area between the constant activity curve (black) and horizontal line $\mathcal{P} = \mathcal{P}^{\rm coexist}$. When the stress-like contribution $\boldsymbol{\Sigma}^{\mathbf{m}}$ is neglected, $I=0$ since the equal-area construction variable is $\mathcal{E}(\rho) = p_{\rm C}$. However, $I>0$ when $\boldsymbol{\Sigma}^{\mathbf{m}}$ is considered and the corresponding coexistence effective pressure (red) should be lower than the coexistence effective pressure predicted by $I=0$ (blue). (b) Coexistence curves for athermal active Brownian spheres (3d) obtained by the mechanical theory with the stress-like contribution of the interparticle forces to the polar order $\boldsymbol{\Sigma}^{\mathbf{m}}$. The parameter $\xi$ represents the magnitude of $\boldsymbol{\Sigma}^{\mathbf{m}}$. The reduced $\mathcal{P}^{\rm coexist}$ resulting from $\boldsymbol{\Sigma}^{\mathbf{m}}$ increases the difference between the coexisting densities. However, this difference, even for extreme values of $\xi$, is not significant.}
	\label{fig:correlation_pressure_binodal}
\end{figure}

Finally, we show that including the stress-like contribution of the interparticle forces to the polar order (i.e. $\boldsymbol{\Sigma}^{\mathbf{m}}$ in the polar order flux [Eq.~\eqref{eq:m_flux_new2}]) broadens the binodals predicted by our mechanical theory, yet only inconsequentially. 
We therefore close our equations by neglecting the stress-like contribution. 
We first note that our first coexistence criterion for MIPS $\mathcal{P}(p_{\rm C}^{\rm liq}) = \mathcal{P}(p_{\rm C}^{\rm gas}) = \mathcal{P}^{\rm coexist}$ is always true regardless of whether we include $\boldsymbol{\Sigma}^{\mathbf{m}}$ or not, as $\boldsymbol{\Sigma}^{\mathbf{m}}$ vanishes in regions of homogeneous density. 
However, since density gradients generate $\boldsymbol{\Sigma}^{\mathbf{m}}$, it can in principle alter the second coexistence criterion, the equal-area construction, by altering $\mathcal{E}$.

From the definition Eq.~\eqref{eq:Sig_m}, it is seen that $\boldsymbol{\Sigma}^{\mathbf{m}}$ represents the correlation between the interaction stress $\boldsymbol{\sigma}^{\rm C}$ and orientation $\mathbf{m}/\rho$. 
We examine the effect of the correlation by considering a constitutive relation $\boldsymbol{\Sigma}^{\mathbf{m}} = \xi \boldsymbol{\sigma}^{\rm C} \mathbf{m} / \rho$. 
Here, a parameter $\xi(>0)$ is introduced in order to investigate effects of the magnitude of this term systematically. The effect of $\boldsymbol{\Sigma}^{\mathbf{m}}$ on the binodal is most clearly seen by considering the following integral:
\begin{equation}
\label{eq:Integral_epsilon_pc}
    I \equiv \int_{p_{\rm C}^{\rm gas}}^{p_{\rm C}^{\rm liq}} \left[ \mathcal{P}(p_{\rm C})  - \mathcal{P}^{\rm coexist} \right] \ d p_{\rm C}  = -\frac{\ell_0}{d-1}  \int_{p_{\rm C}^{\rm gas}}^{p_{\rm C}^{\rm liq}} \frac{d \Sigma^m_{zzz}}{dz} \ d p_{\rm C} \ . 
\end{equation}
When the correlation term is neglected, the integral $I$ trivially vanishes and the corresponding equal-area construction variable is $\mathcal{E}(\rho) = p_{\rm C}$ as discussed in the main text. With the simple model $\boldsymbol{\Sigma}^{\mathbf{m}} = \xi \boldsymbol{\sigma}^{\rm C} \mathbf{m} / \rho$, we find that:
\begin{equation}
\label{eq:integral_with_model}
    I = \frac{\xi \tau_R}{2(d-1)\zeta} \int_{p_{\rm C}^{\rm gas}}^{p_{\rm C}^{\rm liq}} \left(\frac{d p_{\rm C}}{d z}\right)^2 \frac{d}{d p_{\rm C}}\left(\frac{p_{\rm C}}{\rho}\right)\ d p_{\rm C} \ . 
\end{equation}

It can be easily seen that the integrand in Eq.~\eqref{eq:integral_with_model} is always non-negative. 
Consequently, the integral $I>0$ and this implies that the coexistence effective pressure $\mathcal{P}^{\rm coexist}$ is reduced upon including $\boldsymbol{\Sigma}^{\mathbf{m}}$ [see Fig.~\ref{fig:correlation_pressure_binodal}(a)]. 
Accordingly, the difference between the coexisting densities $\rho^{\rm liq}$ and $\rho^{\rm gas}$ increases as shown in Fig.~\ref{fig:correlation_pressure_binodal}(b). 
This is because $\rho^{\rm gas}$ decreases with $\mathcal{P}^{\rm coexist}$ more rapidly than $\rho^{\rm liq}$ does due to the larger $d p_{\rm C} / d \rho$ in the liquid phase. 
Figure~\ref{fig:correlation_pressure_binodal}(b) shows that the coexistence curve is not modified significantly by $\boldsymbol{\Sigma}^{\mathbf{m}}$ even when its magnitude ($\xi$) is large, which allows us to close equations by discarding the term.

\begin{figure}[t]
	\centering
	\includegraphics[width=.95\textwidth]{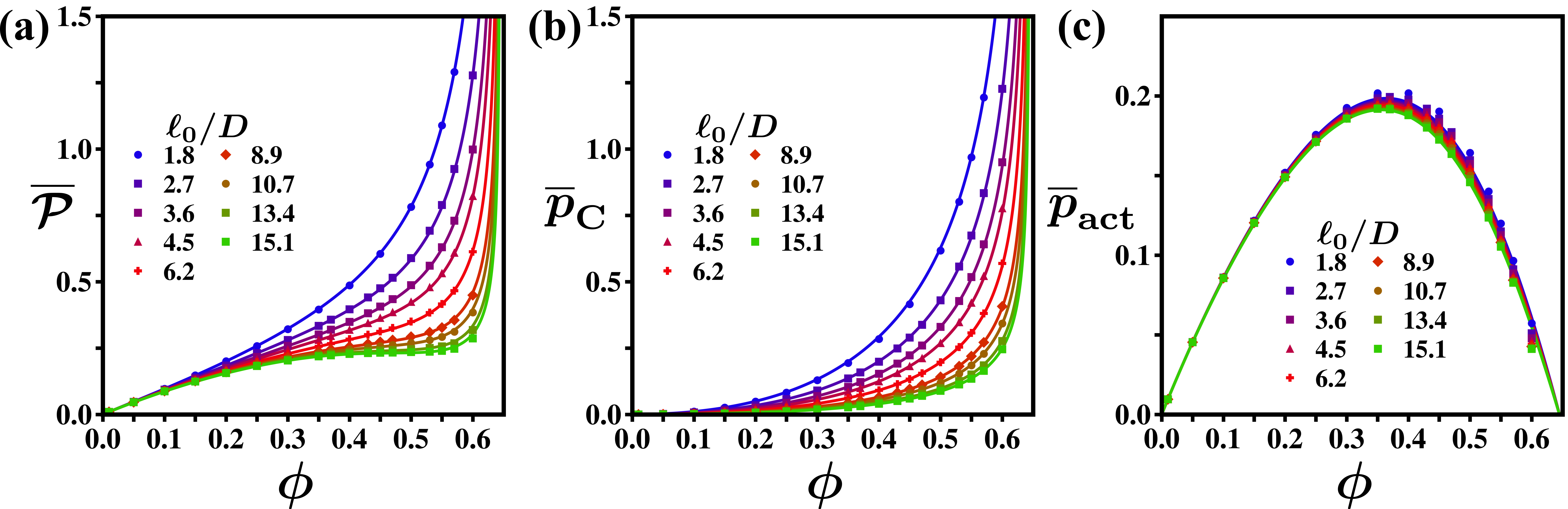}
	\caption{Comparison between simulation data for active Brownian spheres and derived equations-of-state [Eq.~\eqref{eq:3d_DOCS_EOS}] as a function of volume fraction $\phi$. We observe excellent agreement with simulation data for the (a) dynamical pressure, (b) interaction pressure, and (c) active pressure for a range of run lengths $\ell_0/D$ approaching the critical activity $\ell_0^c$. All pressures have been made dimensionless using the active energy scale $\zeta U_0 \ell_0/D^3$ to highlight the collapse and linear behavior of $\bar{p}_{\rm act}$ and $\bar{\mathcal{P}}$ at low volume fractions.}
	\label{fig:total_pressure_3d}
\end{figure}

With our closures, only two quantities are required to describe athermal ABP phase coexistence: the effective active speed $U_0\overline{U}(\ell_0/D, \phi)$ (or equivalently the active pressure) and the conservative interaction pressure $p_{\rm C}(\ell_0/D, \phi)$. 
Accurate equations-of-state in 2d were developed in Ref.~\cite{Mallory2021}, and a detailed derivation of those expressions and a comparison to simulation data can be found in the main text and supplementary material of that work.
A similar procedure to that utilized in Ref.~\cite{Mallory2021} can be used to obtain equations-of-state for active Brownian spheres in three dimensions. The functional form of these expressions are:
\begin{subequations} \label{eq:3d_DOCS_EOS}
\begin{align}
        \frac{p_{\rm act}}{\zeta U_0 / (\pi D^2)} &= \phi \left(\frac{\ell_0}{D}\right) \overline{U} = \phi \left(\frac{\ell_0}{D}\right) \left[1 + \left(1 - \exp\left[-2^{7/6} \left(\frac{\ell_0}{D}\right)\right] \right)\frac{\phi}{1 - \phi / \phi_{\rm max}} \right]^{-1} \ , \label{eq:3d_DOCS_EOS_pi_swim} \\
        \frac{p_{\rm C}}{\zeta U_0 / (\pi D^2)} &=  6\times 2^{-7/6} \frac{\phi^2}{\sqrt{1-\phi/\phi_{\rm max}}} \ , \label{eq:3d_DOCS_EOS_pi_c}
\end{align}
\end{subequations}
where $\phi_{\rm max} = 0.645$ is the maximum random packing fraction achieved in 3d from the simulations, $D = 2^{1/6}\sigma_{\rm LJ}$ is the hard-sphere-like diameter, and $\sigma_{\rm LJ}$ is the Lennard-Jones diameter.
Figure~\ref{fig:total_pressure_3d} provides a comparison between simulation data obtained using Brownian dynamics and the equations-of-state above.

\section{Mechanical Derivation of the Spinodal Condition}
The primary focus of the main text is to obtain the coexistence criteria for mechanically determining the binodal without invoking thermodynamic arguments.
Here, we show that the spinodal -- the region of a phase diagram in which a homogeneous density profile is unstable -- can also be determined mechanically without invoking thermodynamic stability arguments. 
The temporal description of a density profile is provided by the continuity equation:
\begin{equation}
\label{eq:sicontinuity}
\frac{\partial \rho}{\partial t} + \boldsymbol{\nabla} \cdot \mathbf{j}^\rho = 0 \ ,
\end{equation}
where we now require an expression for the number density flux $\mathbf{j}^\rho(\mathbf{x}; t) = \rho(\mathbf{x}; t)\mathbf{u}(\mathbf{x}, t)$ which follows from conservation of linear momentum:
\begin{equation}
\label{eq:silinearmomentum}
\frac{\partial (m \mathbf{j}_\rho)}{\partial t} + \boldsymbol{\nabla} \cdot \left (m\mathbf{j}^\rho \mathbf{j}^\rho/\rho \right) = \boldsymbol{\nabla}\cdot \boldsymbol{\sigma} + \mathbf{b} \ , 
\end{equation}
where $m$ is the particle mass.
Let us consider a \textit{passive} system initially at rest $\mathbf{u}(\mathbf{x}, t_0) = \mathbf{0}$ with an initially homogeneous density profile $\rho(\mathbf{x}, t_0) = \rho_0$.
We now consider small perturbations to the density and velocity fields such that $\rho(\mathbf{x}) = \rho_0 + \delta \rho (\mathbf{x})$ and $\mathbf{u}(\mathbf{x}) = \boldsymbol{\delta} \mathbf{u}(\mathbf{x})$.
Substituting the perturbed density and velocity fields into Eq.~\eqref{eq:sicontinuity} and~\eqref{eq:silinearmomentum} and retaining only terms linear in these perturbations results in:
\begin{subequations}
\begin{equation}
\label{eq:sicontinuityperturb}
\frac{\partial \delta\rho}{\partial t} + \rho_0\boldsymbol{\nabla} \cdot \boldsymbol{\delta} \mathbf{u}= 0 \ ,
\end{equation}
\begin{equation}
\label{eq:silinearmomentumperturb}
m\rho_0\frac{\partial \boldsymbol{\delta} \mathbf{u}}{\partial t}  = \boldsymbol{\nabla}\cdot \boldsymbol{\sigma} \ ,
\end{equation}
\end{subequations}
where we have, for now, neglected body forces.
We require an expression for $\boldsymbol{\sigma}$ to describe the evolution of the density perturbations.
As our focus will be on the behavior of long wavelength perturbations, we omit the spatial gradient terms (e.g., the Korteweg stress or viscous stresses), resulting in $\boldsymbol{\sigma} = -p(\rho)\mathbf{I}$.
The divergence of the stress can now be expressed as $\boldsymbol{\nabla} \cdot \boldsymbol{\sigma} = -(\partial p / \partial \rho) \boldsymbol{\nabla} \delta\rho$ where the compressibility $(\partial p / \partial \rho)$ is to be evaluated at $\rho_0$. 
Differentiating Eq.~\eqref{eq:sicontinuityperturb} with respect to time and substituting in Eq.~\eqref{eq:silinearmomentumperturb} we arrive at:
\begin{equation}
\label{eq:waverealspace}
\frac{\partial^2 \delta\rho}{\partial t^2} = \frac{1}{m}\left (\frac{\partial p}{\partial \rho} \right)_{\rho = \rho_0} \nabla^2 \delta\rho \ ,
\end{equation}
which we recognize as a wave equation for $\delta\rho$ with the wave speed $c$ given by $c ^2 = (\partial p / \partial \rho)/m$. 
Spatially Fourier transforming Eq.~\eqref{eq:waverealspace} we arrive at:
\begin{equation}
\label{eq:wavefspace}
\frac{\partial^2 \delta\rho_k}{\partial t^2} = -(ck)^2 \delta\rho_k \ ,
\end{equation}
where $k$ is the magnitude of the wavevector $\mathbf{k}$ and $\delta \rho_k(\mathbf{k}, t)$ is the Fourier-transformed density perturbation. 
Equation~\eqref{eq:wavefspace} admits a plane-wave solution with:
\begin{equation}
\label{eq:wavefspacesolution}
\delta\rho_k = A_k\exp[-ickt] + B_k\exp[ickt],
\end{equation}
where $A_k$ and $B_k$ are to-be-determined constants.
Clearly, if $c$ is imaginary, the Fourier modes of the density perturbations will grow in time and a homogeneous density is linearly unstable.
This condition only occurs when $(\partial p / \partial \rho) < 0$, thus recovering the same spinodal condition as expected from thermodynamic stability while using only mechanical arguments.

In the case of driven systems, internal body forces $\mathbf{b}(\mathbf{x},t)$ may be generated. 
These internal body forces often have their own evolution equations.
However, there is often a separation of timescales between the relaxation dynamics of the density field and the dynamics of active body forces.
Indeed, in the case of ABPs, the body force arises due to the polarization density of the active force which relaxes on a timescale proportional $\tau_R$ [see Eq.~\eqref{eq:polarization_evolution}].
With a separation of timescales (e.g.,~for timescales large compared to $\tau_R$ in the case of ABPs), we can ignore the dynamics of the body force and, just as in the case of statics (see main text), define an effective stress $\mathbf{\Sigma}$ which incorporates the effects of the body force (for ABPs, $\boldsymbol{\Sigma} = \boldsymbol{\sigma}^{\rm C} + \boldsymbol{\sigma}^{\rm act}$).
Provided that the dynamics of the body force permit the use of this dynamic stress in nonstationary conditions, the analysis applied above for passive systems can be repeated and will result in the spinodal condition of $(\partial \mathcal{P} / \partial \rho) < 0$ with the dynamic pressure ($\boldsymbol{\Sigma} = -\mathcal{P}\mathbf{I}$) now playing a determining role.

In unsteady conditions, the active force density is not the only body force in our model. 
The drag force density also acts as a body force, altering the equation-of-motion of the velocity field from Eq.~\eqref{eq:silinearmomentumperturb} to:
\begin{equation}
\label{eq:silinearmomentumperturbdrag}
m\rho_0\frac{\partial \boldsymbol{\delta} \mathbf{u}}{\partial t}  = \boldsymbol{\nabla}\cdot \boldsymbol{\Sigma} -\zeta\rho_0\boldsymbol{\delta}\mathbf{u}\ ,
\end{equation}
where $\zeta$ is the translational drag coefficient and we now use the dynamic stress.
We can again take a time derivative of Eq.~\eqref{eq:sicontinuityperturb} and now, using Eq.~\eqref{eq:silinearmomentumperturbdrag}, obtain:
\begin{equation}
\label{eq:telegraphrealspace}
\frac{\partial^2 \delta\rho}{\partial t^2} -  \frac{1}{\tau_p}\frac{\partial \delta \rho}{\partial t}= \frac{1}{m}\left (\frac{\partial \mathcal{P}}{\partial \rho} \right)_{\rho = \rho_0} \nabla^2 \delta\rho \ ,
\end{equation}
where $\tau_p = m/\zeta$ is the momentum relaxation time. 
Equation~\eqref{eq:telegraphrealspace} is a telegraph equation. 
Considering timescales much larger than the momentum relaxation time (i.e., $\tau_p \rightarrow 0)$ results in a diffusion equation:
\begin{equation}
\label{eq:diffusionrealspace}
\frac{\partial \delta \rho}{\partial t}= \frac{1}{\zeta}\left (\frac{\partial \mathcal{P}}{\partial \rho} \right)_{\rho = \rho_0} \nabla^2 \delta\rho \ .
\end{equation}
We identify the diffusion coefficient as $D = (\partial \mathcal{P} / \partial \rho)/\zeta$. 
The Fourier space diffusion equation follows as:
\begin{equation}
\label{eq:diffusionfspace}
\frac{\partial \delta \rho_k}{\partial t}= -Dk^2 \delta\rho \ ,
\end{equation}
with a solution of:
\begin{equation}
\label{eq:diffusionfspacesolution}
\delta\rho_k = A_k\exp[-Dk^2t] \ .
\end{equation}
Thus, density perturbations will be linearly unstable only when $D<0$,  again resulting in the same spinodal condition of $(\partial \mathcal{P} / \partial \rho) < 0$.
Finally, we note that it is also straightforward to show that the same spinodal condition is also recovered using Eq.~\eqref{eq:telegraphrealspace} without taking the overdamped limit.

{\color{black} 
\section{Derivation of Mechanical Force Balance by Using Thermodynamic Variational Principle}
We present steps for deriving the mechanical force balance equation (Eq.~(6) in the main text) by using the thermodynamic variational principle. 
Let us consider the free energy functional presented in Eq.~(5) of the main text: $\mathcal{F}[\rho] = \int_V \left[ f  + \rho \mathcal{U}^{\rm ext} + \kappa \left| \boldsymbol{\nabla}\rho \right|^2 / 2 \right]  \ d\mathbf{x}$, where $f(\rho)$ is the mean-field free energy density, $\kappa(\rho)$ is a (positive) coefficient such that the square-gradient term penalizes density gradients~\cite{Cahn1958}, and $\mathcal{U}^{\rm ext}(\mathbf{x})$ represents all externally applied potential fields. 
At constant $(N,V,T)$, the free energy is minimized with respect to any unconstrained internal variables (in this case the density field) when:
\begin{equation}
\label{eq:extremum_condition}
\frac{\delta F}{\delta \rho} =  \frac{\partial f}{\partial \rho} + \mathcal{U}^{\rm ext} - \frac{1}{2} \frac{\partial \kappa}{\partial \rho} \left| \boldsymbol{\nabla}\rho \right|^2 - \kappa  \boldsymbol{\nabla}^2 \rho = \mu^{\rm coexist}\ ({\rm constant})\ ,
\end{equation}
where the constant (coexistence) chemical potential $\mu^{\rm coexist}$ enters as a Lagrange multiplier enforcing the constraint $\int_V \rho\ d\mathbf{x} = N$. 
Equation~\eqref{eq:extremum_condition} simply states that chemical potential should be constant throughout the volume at equilibrium. 
We now cast this thermodynamic equilibrium condition in the form of a mechanical force balance. 
To do so, we recognize that the external force density per volume, i.e. $- \rho \boldsymbol{\nabla}\mathcal{U}^{\rm ext}$, should be present in the final mechanical force balance. 
This motivates us to multiply Eq.~\eqref{eq:extremum_condition} with $\boldsymbol{\nabla}\rho$ and we obtain:
\begin{equation}
\label{eq:pre_mechanical_force_bal}
- \rho \boldsymbol{\nabla} \mathcal{U}^{\rm ext} 
+ \boldsymbol{\nabla} \cdot
\left[
\left(
f  + \rho \mathcal{U}^{\rm ext} - \mu \rho + \frac{1}{2} \kappa \left| \boldsymbol{\nabla}\rho \right|^2 
\right)\mathbf{I} - \kappa  \boldsymbol{\nabla} \rho \boldsymbol{\nabla} \rho
\right] = \mathbf{0} \ .
\end{equation}
Using Eq.~\eqref{eq:extremum_condition} and a thermodynamic relation $p(\rho) = -f(\rho) + \rho \partial f/\partial\rho$, we finally obtain:
\begin{equation}
\label{eq:mechanical_force_bal}
- \rho \boldsymbol{\nabla} \mathcal{U}^{\rm ext} 
+ \boldsymbol{\nabla} \cdot
\left[
- p\mathbf{I} + \left(\frac{1}{2} \frac{\partial ( \kappa \rho )}{\partial \rho} \left|\boldsymbol{\nabla} \rho \right|^2 + \kappa \rho \boldsymbol{\nabla}^2 \rho \right) \mathbf{I} 
    - \kappa \boldsymbol{\nabla}\rho \boldsymbol{\nabla}\rho
\right]  = \mathbf{0}  \ .
\end{equation}
Equation~\eqref{eq:mechanical_force_bal} is the mechanical force balance in equilibrium -- the externally imposed body force $\mathbf{b}  = -\rho \boldsymbol{\nabla} \mathcal{U}^{\rm ext}$ is balanced by the divergence of stress $\boldsymbol{\sigma} = - p\mathbf{I} + \left(\partial ( \kappa \rho )/\partial \rho \left|\boldsymbol{\nabla} \rho \right|^2 / 2+ \kappa \rho \boldsymbol{\nabla}^2 \rho \right) \mathbf{I} - \kappa \boldsymbol{\nabla}\rho \boldsymbol{\nabla}\rho$.
}